\documentclass[dvips,11pt,a4paper]{article}
\usepackage[dvips,colorlinks,bookmarksopen,bookmarksnumbered,citecolor=red,urlcolor=blue]{hyperref}

\usepackage[T1]{fontenc}         
\usepackage{graphicx}
\usepackage{amsmath}
\usepackage{bm}
\usepackage{tikz}
\usepackage{color}
\usepackage{helvet}
\usepackage{courier}
\usepackage{makeidx}
\usepackage{footmisc}
\usepackage{type1cm}         
\usepackage{textcomp}
\usepackage{subfig}
\usepackage{enumitem,xcolor}    

\DeclareCaptionLabelSeparator{colon}{. }
\usepackage[labelfont={bf},labelsep={colon}]{caption}

\makeindex

\DeclareMathAlphabet\mathbfcal{OMS}{cmsy}{b}{n}

\begin{document}
\thispagestyle{empty}

\begin{center}
\textcolor{blue}{ \Large  \bf Unified discrete approach of acceleration conservation } \\
\vspace{3.mm}
{\bf Jean-Paul Caltagirone } \\
\vspace{3.mm}
{ \small Universit{\'e} de Bordeaux  \\
   Institut de M{\'e}canique et d'Ing{\'e}ni{\'e}rie \\
   Département TREFLE, UMR CNRS n° 5295\\
  16 Avenue Pey-Berland, 33607 Pessac Cedex  \\
\textcolor{blue}{\texttt{ calta@ipb.fr }  } }
\end{center}

\vspace{2.mm}
\small
{\bf Abstract}

Discrete mechanics is used to present fluid mechanics, fluid-structure interactions, electromagnetism and optical physics in a coherent theoretical and numerical approach. Acceleration considered as an absolute quantity is written as a sum of two terms, {\it i.e.} an irrotational and a divergence-free component corresponding to a formal Hodge-Helmholtz decomposition. The variables of this equation of discrete motion are only the scalar and vector potential of the acceleration, whatever the physical field. These, like the physical properties, are only expressed as a function of two fundamental units, namely a length and a time.

The numerical methodology associated with this equation of motion is based on discrete operators, gradient, divergence, primal and dual curl applied to the velocity components of the primal geometric topology. Some solutions resulting from simulations carried out in each domain make it possible to find the results obtained from the Navier-Stokes, Navier-Lam{\'e} and Maxwell equations and to show the coherence of the proposed unified approach.

\vspace{2.mm}



\section{Introduction}

The Navier-Stokes, Navier-Lam{\'e} and Maxwell equations are fully representative of physical phenomena in the fields of fluid mechanics, solid mechanics and electromagnetism. Continuum mechanics \cite{Tru74}, \cite{Tru92} was supposed to unify mechanics but two centuries after their derivation, it turns out that its (Eulerian or Lagrangian) formulations are different; velocity is the variable in fluid media whereas it is displacement in solid media.
Moreover, the mass conservation equation is closely associated with the Navier-Stokes equation which cannot work without the law of mass conservation, while the Navier-Lam{\'e} equation is autonomous. 
The Lam{\'e} coefficients, compression and shear modulus are perfectly measurable for a solid, whereas compression viscosity does not have an obvious experimental value other than that given by the Stokes law, $\lambda = - 2/3 \: \mu$. This law is wrong \cite {Gad95b}, \cite {Raj13}. The difficulty is one of the {\it artifacts} derived directly from the concept of continuum \cite {Cal19a}.

Along with mechanics, electromagnetism is the main domain in which the problem of the unification of certain laws of physics has arisen. It was successfully solved by J.C. Maxwell, based on the laws of Faraday, Ampère, Thomson, Gauss and Joule, etc., by introducing the fundamental notion of dynamics \cite{Max65}. Numerous attempts at analogies between fluid mechanics and electromagnetism have been made without unification being proved. From a numerical point of view, the aim is to design dedicated methodologies that make it possible to federate these two domains, such as in the works of A. Bossavit \cite{Bos03} for finite elements. 

The contribution of E. Tonti \cite{Ton02}, \cite{Ton13} illuminates the problem differently by presenting an algebraic formulation of different physical laws without resorting to a process of discretization of the differential equations. This concept differs fundamentally from the classical approach in which the formulation of differential equations is followed by their discretization. Although the case of fluid mechanics has its particularities, this concept makes it possible to establish direct analogies between the different quantities in most domains of the fields of physics. Tonti considers that computational physics is richer than computational mathematics; it could be added that discrete physics possesses properties that continuum physics does not.

These different attempts do not focus on the fundamental differences between the notion of a continuous medium highlighted in mechanics and the inherently undulatory and particulate nature of electromagnetism. Important differences appear: in principle, the equation of motion is not an equation of wave mechanics and the Maxwell equations do not include the notion of inertia. Basically, any equation unifying these two domains of physics requires an inertial wave formulation to be proposed. The Navier-Stokes equations for fluids, Navier-Lam{\'e} for solids and Maxwell equation have been validated by a multitude of experimental observations due, of course, to the satisfaction of the constitutive laws elaborated over time. If an equation allowing the unification of the laws of physics can be elaborated, in principle it must make it possible to find all the observations and results obtained previously. The point of view developed here is that, even though the results of the Navier-Stokes and Maxwell equations are in agreement with the observations, the physical models themselves exhibit artifacts, flaws or inadequacies.

The potential unification of the laws of physics requires the representativeness of the equation corresponding to the whole spectrum of frequencies. In the course of time, undulatory mechanics has been transformed into quantum mechanics, see for example L. Brillouin \cite{Bri26}. Wave optics, meanwhile, is now called physical optics and, under the impetus of N. Bohr, R. Feynman and A. Einstein, the union of two aspects is defined by wave-particle duality.

Some aspects of electromagnetism and physical optics are found in certain behaviors of new synthetic materials, such as the observation of interference fringes, polarization phenomena, and chirality, etc. For example, the metamaterials developed by T. Frenzel \cite{Fre17}, \cite{Cou17} have the particularity of rotating when they are subjected to compression. The field opened up by the interpretation of certain optical phenomena applied to the development of materials that possess similar properties is very promising, especially with the use of 3D printers. However, the present contribution will relate only to the fundamental concepts, while the associated physical properties will also allow us to account for the complex behaviors that are observed or sought.

The objective here is not to find any analogy, scaling or dimensionless unitary form between these phenomena, but to propose a truly unified formulation composed of a single equation system associated with common variables. The extension of discrete mechanics to electromagnetism made it possible to find the main properties of Maxwell's equations on electric and magnetic fields by the degeneracy of the discrete equation. This is formulated as a Hodge-Helmholtz decomposition of the acceleration considered in discrete mechanics as an absolute quantity. Velocity is a relative secondary quantity which must be cleaned of its constant values corresponding to uniform motions of translation and rotation. Discarding the notion of continuous medium \cite{Cal19a} leads us also to discard the notion of a global reference frame in favor of a description based on a local reference frame.

The equation of motion is thus the sum of a divergence-free term and an irrotational one, each of which is attached to a scalar potential $\phi$ and a vector potential $\bm \psi$ respectively. The solution to a problem of mechanics or electromagnetism is therefore associated only with the variables $(\mathbf V, \phi, \bm \psi)$. All other quantities can therefore be expressed from the generic variables, using a one-to-one correspondence with the usual variables. It turns out that all the quantities, variables and physical characteristics are expressed with only two fundamental units, those of a length and a time.

After a presentation of the basic concepts of discrete mechanics, examples of numerical simulations from classical test cases in fluid mechanics, solids, electromagnetism and optical physics demonstrate the versatility and validity of the proposed discrete approach.

\section{Discrete formulation}

\subsection{Bases of discrete medium}

The vision proposed here is based on the existence of a single law for all the areas of physics mentioned above.
The equivalence principle of Galileo (Weak Equivalence Principle) is now experimentally verified on the one part of $10^{15}$, see for example C.M. Will \cite{Wil18}. If $\bm \gamma$ is the proper acceleration of a material medium or a particle and $\mathbf g$ is the acceleration of gravity this principle allows to write:
\begin{eqnarray}
\left\{
\begin{array}{llllll}
\displaystyle{ m_0 \: \bm \gamma_1 =  m_0 \: \mathbf g_1 = m_0 \: \nabla \left( \frac{\mathcal{G} \: M_1}{r} \right) } \\  \\
\displaystyle{ m_0 \: \bm \gamma_2 =  m_0 \: \mathbf g_2 = m_0 \: \nabla \left( \frac{\mathcal{G} \: M_2}{r} \right) } \\  \\
\displaystyle{ m_0 \: \bm \gamma = m_0 \: \left( \bm \gamma_1 + \bm \gamma_2 \right) =  m_0 \: \left( \mathbf g_1 + \mathbf g_2 \right) } 
\end{array}
\right.
\label{gravit}
\end{eqnarray}
where $m_0$ is the rest mass of the particle and $\mathcal G \: M / r$ the gravitational potential where $\mathcal G$ is the gravitational constant and $r$ is the distance between the particle and the object of mass $M$.

According to WEP the mass of the particle $m_0$ does not appear in the fundamental principle of  dynamics and the equalities (\ref{gravit}) lead to:
\begin{eqnarray}
\displaystyle{ \bm \gamma   = \mathbf g_1 + \mathbf g_2  } 
\label{princip}
\end{eqnarray}

Acceleration is an additive quantity in Newtonian mechanics for gravitation.
Can the law (\ref{princip}) be generalized to other physical effects than gravitation? Several arguments can be advanced:
\begin{itemize}
\item All S.I. units of the physical quantities that contain the  mass are only in order one, so it is possible to define equivalent quantities per unit mass;
\item the physical modeling of  all the physical phenomena of the field theory (viscosity, inertia, dissipation, compression, gravitation, ...) can be described using the only two quantities, a length and a time;
\item the discrete motion equation established without the mass gives identical results to the classical approach including for problems with variable densities.
\end{itemize}

In the context of a local coordinate system additivity accelerations remains the rule. An example: a satellite  in geostationary orbit around the Earth is in mechanical equilibrium; the acceleration of the satellite in its own frame of reference is null $\bm \gamma = \mathbf g_1 + \mathbf g_2 = 0$ where $\mathbf g_1 = \nabla (\mathcal{G} \: M / r)$ is the acceleration of earth gravity and $\mathbf g_2 = \nabla ( | \mathbf V |^2 / 2)$ is the centrifugal acceleration (or inertia) due to its own rotation. This equilibrium does not depend on the mass of the satellite.

In discrete mechanics an observer on  local frame of reference is not aware of his own mass and all the interactions with the neighborhood  are of cause and effect. 
This observer in mechanical equilibrium defined by a null acceleration can not detect if it is in uniform translation or in uniform rotation.

The principle of relativity and that of the equivalence of inertial and gravitational masses (WEP)  suggest the existence of an invariant absolute variable independent of any frame of reference. This quantity is the acceleration $\bm \gamma$ taken by a medium or particle under the influence of the acceleration imposed on it. Given the principle of equivalence, the generic law is written: 
\begin{eqnarray}
\displaystyle{ \bm \gamma = \mathbf g  } 
\label{loiphys}
\end{eqnarray}

 This law conforms to Newton's second law $m \: \bm \gamma = \mathbf F$, but here the vector $\mathbf g$ is the set of forces per unit mass applied to the medium or particle. The law (\ref{loiphys}) expresses the conservation of acceleration. It is the only physical quantity which satisfies the mathematical vector addition. 
The law $\bm \gamma = \mathbf g$ expresses the acceleration conservation.

The mass considered in the fundamental law of dynamics in classical or Newtonian mechanics is the constant rest mass $m = m_0$. In the general case, the mass or density depends on velocity. In special and general relativity, the equation is written $d (m_r \: \mathbf V) / dt = \mathbf F$ where mass $m_r$ is relativistic $m = m_r = \gamma \: m_0 = m_0 / \sqrt{1 - v ^ 2 / c_0 ^ 2}$. In continuum mechanics, the conservation of mass leads to a local motion equation of the form $\rho \: d \mathbf V / dt = \rho \: \mathbf g $.
Discrete mechanics, where $\bm \gamma = \mathbf g$, corresponds to a third vision that is not in conflict with the first two and is an alternative approach.

 This law (\ref{loiphys}) is the cornerstone of the discrete mechanics developed in recent years \cite{Cal15}.
The basic assumptions of discrete mechanics are simply recalled here:
\begin{itemize}
\item the acceleration of a particle or a medium is an absolute quantity in a local reference frame;
\item velocity is not limited, celerity and velocity are two disjointed notions; the first is a property of the medium (matter, vacuum) and the second is a relative quantity which can be accumulated;
\item the equivalence of gravitational and inertial masses and relativity are two intangible principles;
\item there is a scalar potential $\phi$ and a vector potential $\bm \psi$ of the same quantity $\bm \gamma$, acceleration;
\item the Hodge-Helmholtz decomposition applies to the acceleration which is broken down into an irrotational component and a solenoidal one;
\item source terms resulting from physical effects, such as inertia, gravity, and capillarity, etc., can be broken down according to this same principle. 
\end{itemize}

Any vector can be broken down into an irrotational part and a divergence-free part:
\begin{eqnarray}
\displaystyle{ \bm \gamma = - \nabla \phi + \nabla \times \bm \psi  } 
\label{dechh}
\end{eqnarray}

 We will adopt the principle that any vector can be written in this form. This decomposition is sometimes presented with a third harmonic term, both divergence-free and irrotational. In fact, this term is closely associated with the uniform overall movements that must disappear from the formulation under the principle of relativity. Velocity is a variable whose absolute value is not required. In discrete mechanics it is considered as a simple Lagrangian upgraded by acceleration $\mathbf V = \mathbf V^o + dt \: \bm \gamma $ where $\mathbf V^o $ is the velocity at moment $t^o$ and $dt$ the elapsed time between two observations of the phenomenon. 

The acceleration is therefore written as the sum of the two terms $\bm \gamma = \bm \gamma_{\phi} + \bm \gamma_{\psi}$, which represent the direct and induced accelerations and which can be modeled according to velocity and transport properties.
Velocity $\mathbf V$ also has two components that are upgraded by the components of the acceleration $\mathbf V_{\phi} = \bm \gamma_{\phi} \: dt$ and $\mathbf V_{\psi} = \bm \gamma_{\psi} \: dt$, which represent the flux over the segment $\Gamma$ and, as the two fields come from orthogonal operators, these velocities do not have any direct interaction.
In fact, it is not possible, in general, to extract the components of velocity $\mathbf V$ by directly applying a local discrete Hodge-Helmholtz decomposition. The result depends closely on the boundary conditions \cite{Bha12}. Fortunately, the $\mathbf V$ field is relative and is only a secondary variable; it is the decomposition of the acceleration $\bm \gamma$, the absolute quantity, which is sought in the form of potentials and this is possible through the fundamental law (\ref{loiphys}).

\subsection{Primal and dual geometric topologies}

\begin{figure}[!ht]
\begin{center}
\includegraphics[width=6.cm,height=4.cm]{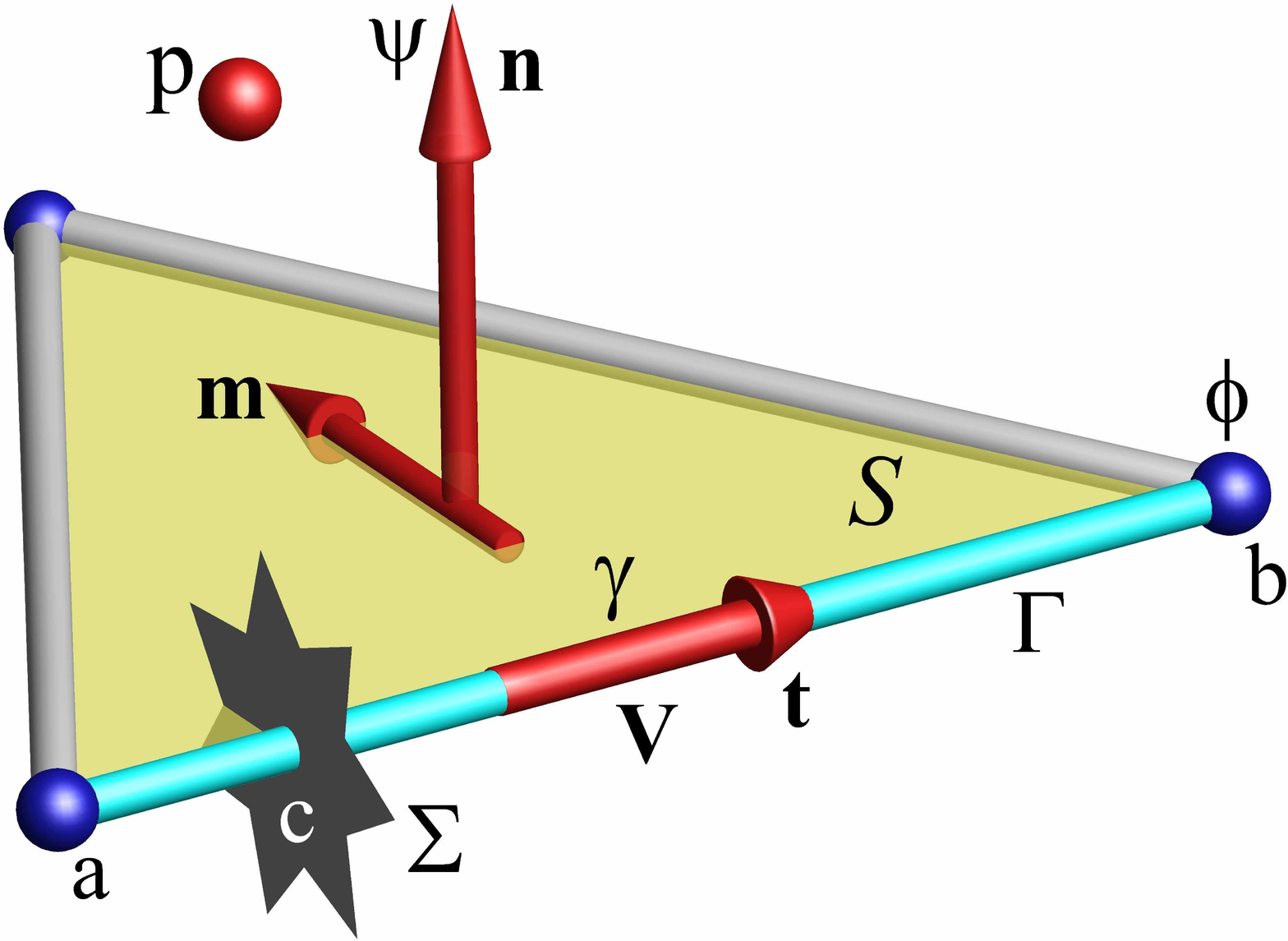}
\hspace{10.mm}
\includegraphics[width=5.6cm,height=6.cm]{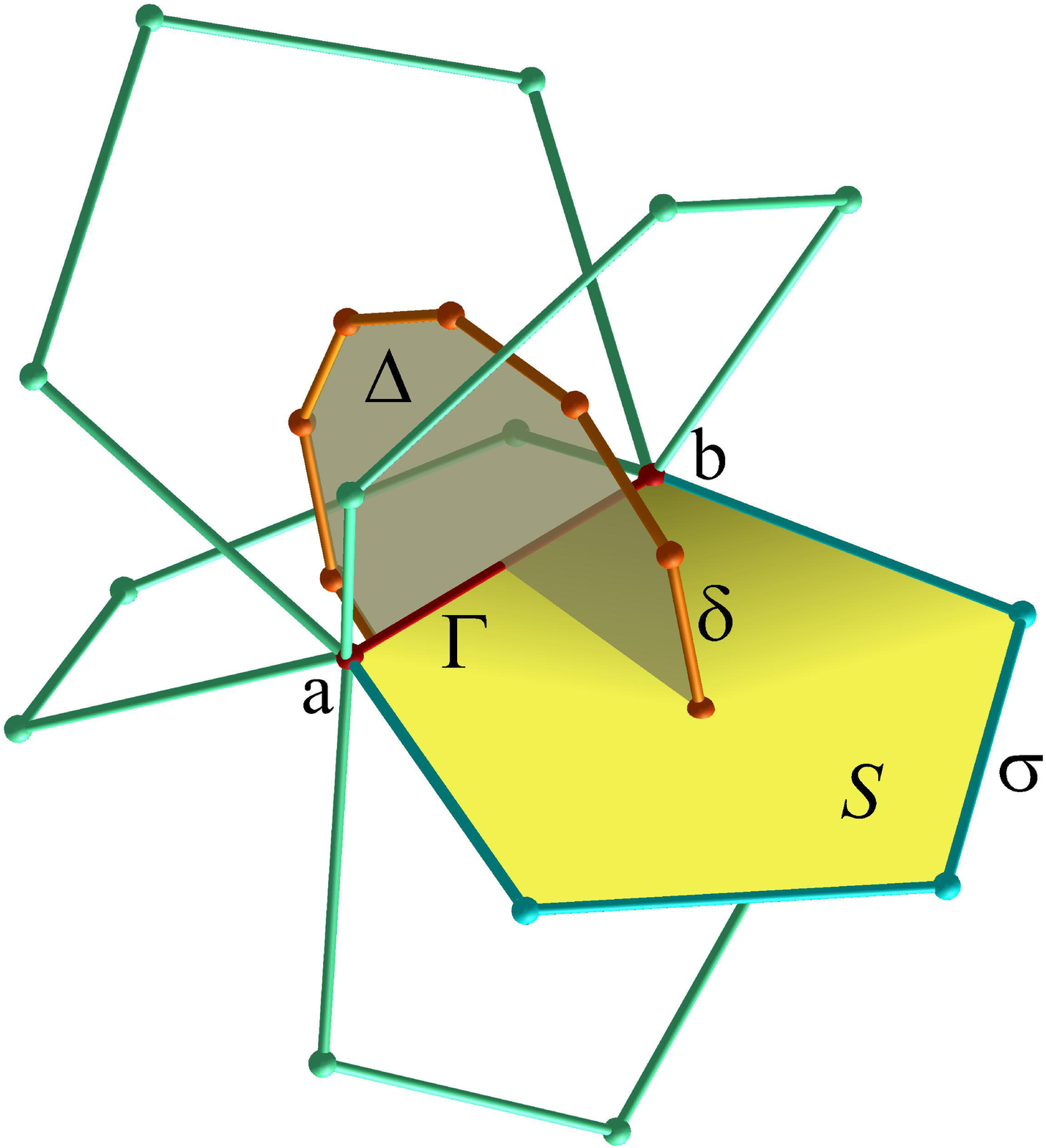}  \\
\vspace{-1.mm}
\hspace{-1.mm} (a) \hspace{60.mm} (b)
\caption{\it (a) Elementary geometrical structure of discrete mechanics in a direct reference frame $(\mathbf m, \mathbf n, \mathbf t)$: three straight edges $\Gamma$ delimited by dots define a planar face $\mathcal S$. The unit normal vectors $\mathbf n$ on the face and the vector carried by $\Gamma$ are orthogonal, $\mathbf t \cdot \mathbf n = 0$. The edge $\Gamma $ can be intercepted by a discontinuity $\Sigma$ located in $c$, between the ends $a$ and $b$ of $\Gamma$. $\phi$ and $\mathbf \Psi$ are the scalar and vector potentials respectively. (b) The virtual machine of motion in Discrete Mechanics: the acceleration of the medium along the edge $\Gamma$ is due to the difference in the scalar potential $\phi$ between the vertices of the edge $[a, b]$ of unit vector $\mathbf t$, and to the circulation action of the vector $\mathbf V$ on the contour of the different primal facets $ \mathcal S $ inducing an acceleration on $\Gamma$. Particle $p$ has a velocity and an acceleration whose projections on the $\Gamma$ edges are named respectively $\mathbf V$ and $\bm \gamma$.  }
\label{discreet}
\end{center}
\end{figure}

The notion of continuum is also abandoned, as well as that of a global frame of reference; there is a local discrete geometric topology represented in the figure (\ref{discreet}), composed of a primal topology and a dual one. The oriented segment $\Gamma$ of unit vector $\mathbf t$ of ends $a$ and $b$ defines the basic element of the primal topology which forms, with two other edges, the planar surface $\mathcal S$ whose unit-oriented normal is $\mathbf n$ such that $\mathbf t \cdot \mathbf n = 0$ (figure \ref{discreet}(a)). The scalar potential $\phi$ is only defined at the ends of the primal topology. Any contact discontinuity or shock wave $\Sigma$ intersects the segment $\Gamma$ at $c$. The normal on the $\mathcal S$ surface is associated with a pseudo-vector $\bm \psi$, such that the rotation of the vector $\mathbf V$ is itself associated with the segment $\Gamma$. Figure (\ref{discreet}(b)) represents the polydual discrete elements composed of primal surfaces $\mathcal S$ in the form of planar polygons; the outline $\delta$ and surface $\Delta$ form the dual topology.

The material medium, a flux of particles or an isolated particle represented by a sphere $ p $ in figure (\ref {discreet}(a)) has a velocity and a spin, but only the components of velocity and momentum are represented on each edge $\Gamma$. Whereas in classical Lagrangian mechanics it is not possible to attribute a rotation of the particle $p$ on itself, the projection of its motion, translation and rotation on $\Gamma$ makes it possible to take electrodynamic interactions into account.

The presentation of the differential operators may differ greatly depending on the degree of formalization of the differential geometry \cite{Mar02}. The succinct and non-exhaustive presentation given here is based on a simple physical approach which allows us to define the operators associated with the switch from one topology to another on the basis of scalar or vectorial information. It should be remembered that although the classic notion of a continuum has been set aside, the material is a geometrical structure where the directions of the edges and of the normal to the surfaces are also preserved at all scales of observation.

The gradient operator applied to a scalar $\phi$, $\nabla \phi$ represents the difference of that scalar over a distance $d$ between the vertices $a$ and $b$ in a given direction. Unlike with the concept of continuum mechanics, the gradient vector defined here has only one component, assigned as a scalar to the edge $\Gamma$. The gradient of a scalar in space has no meaning - it is an illegal operation in discrete mechanics. The gradient is calculated solely on a bipoint $[a b]$ linked by an edge. The primal curl of a vector $\mathbf{W}$, $\nabla \times \mathbf{W}$ is associated with the circulation of its components $\mathbf{V}$ over all the edges $\Gamma$ constituting the primal topology. It is represented by a vector $\mathbf{n}$ orthogonal to the planar primal surface. This apparent restriction disappears as the surface area $ds$ tends toward zero; however, it will remain a condition for the application of the theorems of differential geometry in the particular context discussed here. The divergence represents the flux of a vector $\mathbf{W}$, $\nabla \cdot \mathbf{W}$ across all the facets of the dual surface. The scalar that is obtained is assigned to the single point inside the dual volume. The flux is calculated on the basis of the components $\mathbf{V}$ on the edges $\Gamma$ of the vector itself. If the vector $\mathbf{W}$ is a curl, calculated as the circulation of another vector on each primal topology, then the divergence will be strictly null. The dual curl of a vector $\mathbf{W}$, $\nabla \times \mathbf{W}$ physically represents the flux of the vector $\mathbf{W}$ across that portion of the dual surface associated with the edge $\Gamma$. This flux is calculated using the circulation of the vector, or rather, of its components, on the boundary delimiting the dual portion which, in general, is not planar. The result of this operation is assigned to the edge $\Gamma$ as a vector or a scalar on an oriented edge.

The circulation of the vector $\mathbf V$ along the contour of the primal surface makes it possible to calculate the primal curl carried by the unit vector $\mathbf n$; the circulation along the contour $\delta$ of the surface $\Delta$ is the dual curl that re-projects the result on the segment $\Gamma$. Note that the number of planar surfaces of normals $\mathbf n$ associated with the segment $\Gamma$ is arbitrary, five in figure (\ref{discreet}(b)). In the context of a continuous medium, the necessity of using a global frame of reference leads us to consider the three components of the vectors, the nine components of a second-order tensor, and so on. The same reasons require the introduction of the four components of space-time, the fourth-order Riemann tensor in relativity theory. The discrete description makes it possible to satisfy the notion of material indifference from the outset and to represent the notion of polarization. 

The primal and dual topologies thus defined make it possible to satisfy two essential properties, $\nabla_h \times \nabla_h \phi = 0$ and $\nabla_h \cdot (\nabla_h \times \bm \psi) = 0$, whatever the topologies based on planar surfaces, polygons or polyhedra, and whatever the regular functions $\phi$ and $\bm \psi$. These conditions are absolutely necessary for a complete Hodge-Helmholtz decomposition applied here to acceleration. Each vector can be decomposed into a solenoidal part and an irrotational part, but the scalar and vectorial potentials are not of the same importance according to the nature of the vector. In electromagnetism, $\phi$ and $\mathbf A$ do not come from the same vector and do not express themselves with the same units. In mechanics, the scalar potential and velocity vector potential do not have a particular physical importance but can be used to project velocity over a field with zero divergence. Only acceleration $\bm \gamma$ and its potentials $\phi$ and $\bm \psi$ have special physical properties.

It should be noted that $\phi^{o}$ and $\bm{\psi}^{o}$ are the stresses at time $t$, where all the forces applied before that instant are ``remembered''. The formalism presented here enables us to take account of the entire history of the medium, i.e. its evolution over time from an initial neutral state. For a given instantaneous state of strain, there may be multiple paths by which that state can be reached, and $(\phi^{o},\bm{\psi}^{o})$ will, alone, contain the whole of the medium\textquoteright s history. It is not helpful to know the local and instantaneous stresses, as these two potentials will have accumulated stresses over time; these quantities are also called ``accumulators'' or ``storage potentials''. These potentials can therefore be used to take account of the behavior of media with continuous memory.

The system of equations (\ref{discrete}) is written using discrete operators that apply to either $\mathbf t$ oriented vectors or to the facet normal $\mathbf n$ without the need to specify the primal or dual nature of the operator. However, at this stage, and in order to remove any ambiguity, the action of the subscripted notation $p$ or $d$ is introduced in the table (\ref{schemtab}).
\begin{table}[!ht]
\begin{center}
\begin{tabular}{|c|c|}   \hline
 $ \nabla_d \left( \phi - dt \: c_l^2 \: \nabla_p \cdot \mathbf V \right) = \mathbf g_l$  & $ \nabla_d \times \left( \bm \psi - dt \: c_t^2 \: \nabla_p \times \mathbf V \right) = \mathbf g_t$  \\\hline
 $ \nabla_p \cdot \left( \nabla_d \times \bm \psi \right) = 0$  & $ \nabla_p \times \left( \nabla_d \: \phi \right) = 0 $  \\\hline
\end{tabular}
\caption{ \it Summary of the action of the discrete operators for the applied acceleration $\mathbf g$ (first line) and fundamental properties of these same operators (second line). The velocity component is denoted $\mathbf V$ and $\phi$ and $\bm \psi$ are the scalar and vector potentials of the acceleration $\bm \gamma $; the primal operator is denoted $\nabla_p$ and the dual operator $\nabla_d$. }
\label{schemtab}
\end{center}
\end{table}

This table summarizes the actions of discrete operators for obtaining the external acceleration $\mathbf g = \mathbf g_l + \mathbf g_t$ in the form of a Hodge-Helmholtz decomposition. For the first line, the external operators re-project the action of the internal operators on the edge $\Gamma$. The second line reproduces the fundamental properties of the continuous medium mimicked by the discrete formulation from the primal and dual operators.

The notions of velocity and celerity specific to this discrete context must be specified. The velocity vector $\mathbf W$ or its components on each oriented edge $\Gamma$ of the primal geometric topology represents the amplitude of velocity of the material medium or of the particle; its material derivative $d \mathbf V / dt$ is acceleration $\bm \gamma$. Celerity $ c $ corresponds to the speed of the wavefront. It is a constant scalar equal to $c$, for exemple to $c_0$ the speed of light in a vacuum; in the general case, this property is a function of other variables. The velocity $\mathbf V$ is constant on the edge $\Gamma$, while the celerity is set to vertex $a$ or $b$.

\subsection{Discrete motion equation} 

The equation of motion is established for all media, fluids, solids or vacuum. Properties can be any, including:
\begin{itemize}
\item the fluids can be compressible or incompressible, Newtonian or complex rheologies, nonlinear, viscoelastic, viscoplastic, with thresholds, etc.;
\item the behaviors of solids can correspond to various constitutive laws, from the elastic solid to complex constitutive laws; the unsteady temporal processing of the model makes it possible to treat cases of large deformations and large displacements while conserving the mass;
\item electromagnetic media can have any properties corresponding to those of a fluid medium, solid or vacuum; persistent states, such as electric charge accumulation, permanent magnetization or hysteresis effects, are implicitly incorporated into the formulation.
\end{itemize}

The derivation of the previously realized equation of motion \cite{Cal19a} leads to the system of generic equations of discrete mechanics:
\begin{eqnarray}
\left\{
\begin{array}{llllll}
\displaystyle{ \bm \gamma = - \nabla \left( \phi^o - dt \: c_l^2 \: \nabla \cdot \mathbf V \right) + \nabla \times \left( \bm \psi^o - dt \: c_t^2 \: \nabla \times \mathbf V \right)  + \mathbf g } \\  \\
\displaystyle{ \phi  = \alpha_l \: \phi^o - dt \: c_l^2 \: \nabla \cdot \mathbf V } \\ \\
\displaystyle{ \bm \psi  = \alpha_t \: \bm \psi^o - dt \: c_t^2 \: \nabla \times \mathbf V } \\ \\
\displaystyle{ \mathbf V = \mathbf V^o + \bm \: \gamma \: dt } 
\end{array}
\right.
\label{discrete}
\end{eqnarray}

The quantity $\phi^o$ is the mechanical equilibrium scalar potential and $\bm \psi^o$ is the equilibrium vector potential. The source term  $\mathbf g$, an acceleration, represents all the other effects that can be applied. The factors $\alpha_l$ and $\alpha_t$ are physical quantities to express the persistence of long-term effects, for example the relaxation time of shear stresses in a fluid medium is of an order of magnitude of $\tau_f \approx 10^{-12} \: s$ and the $\alpha_t$ factor can legitimately be set to zero for common applications. In continuum mechanics terminology, the quantity $\nabla \phi$ is a polar vector just like $\nabla \times \bm \psi$, whereas $\bm \psi$ is an axial vector or pseudo-vector. In the absence of any movement of the particle or medium and in the presence of a source term $\mathbf g$, the equation becomes $- \nabla \phi^o + \nabla \bm \psi^o + \mathbf g = 0$. They express the persistence of a physical phenomenon, such as electric potential and permanent magnetization in electromagnetism or pressure and shear stress in a solid medium in mechanics.

A drag representing the effects of viscous friction on small scales $- \kappa \: \mathbf V$ can be added to the equation; in fluid mechanics, this term is the drag of Darcy $- (\nu_f / K) \: \mathbf V $ where $\nu_f$ is the kinematic viscosity of the fluid and $K$ is the permeability of the porous medium. In electromagnetism, this term enables modeling of the reduction in the acceleration of electric charges.

The longitudinal $c_l$ and transverse $c_t$ celerities are intrinsic properties of the medium, matter or vacuum; these quantities depend on multiple variables, but they will be assumed to be simply known in space and time.  The definition of these celerities also depends on the physical phenomenon studied, electromagnetism, fluid mechanics, etc. Velocity $\mathbf V$ is a quantity that is totally independent of celerity, the two notions being strictly disjoint.
Velocity is a secondary variable, a lagrangian, which is updated from the acceleration and elapsed-time $dt$ between two observations of the physical system, $\mathbf V = \mathbf V^o + dt \: \bm \gamma$.

The system (\ref{discrete}) is composed of a vectorial equation whose variable is $\mathbf V$ and three updates of potentials and velocity. The acceleration $\bm \gamma$ can be replaced by the material derivative definition $\bm \gamma = d \mathbf V / dt$ to give an implicit law. The material derivative itself can be replaced by expressing the terms of inertia $\bm \gamma_i = \partial \mathbf V / \partial t - \nabla \times (| \mathbf V |^2/2 \: \mathbf n) + \nabla (| \mathbf V |^2/2) $. The form of the inertial terms discussed in \cite{Cal19a} is applicable whatever the medium considered, fluid, solid or vacuum.
Similarly, all possible source terms $\mathbf g$ are considered accelerations and thus decomposed in Hodge-Helmholtz form. This is the case of gravity where the scalar potential is $\phi_g = \mathcal G \: M / r$, thus giving the two contributions of gravitational acceleration $\mathbf g_g = - \nabla \phi_g + \nabla \times \phi_g \: \mathbf n$; another example is that of capillary acceleration which is written $\mathbf g_c = - \nabla \phi_c + \nabla \times \phi_c \: \mathbf n $ with $\phi_c = \sigma \: \kappa$ where $\sigma$ is the surface tension per unit mass and $\kappa$ the longitudinal or transverse curvature.

These contributions of potentials $\phi_i$, $\phi_g$ and $\phi_c$ can be summarized by accelerations: 
\begin{eqnarray}
\left\{
\begin{array}{llllll}
\displaystyle{ \bm \gamma_i =  \nabla  \left( \frac{| \mathbf V |^2 }{2}  \right) - \nabla \times \left( \frac{| \mathbf V |^2 }{2} \: \mathbf n \right)  } \\  \\
\displaystyle{ \bm \gamma_g =  \nabla  \left( \frac{ \mathcal G \: M }{ r} \right) - \nabla \times \left( \frac{ \mathcal G \: M }{ r} \: \mathbf n \right)  } \\  \\
\displaystyle{ \bm \gamma_c =  \nabla  \left(  \sigma_l \: \kappa_l  \right) - \nabla \times \left( \sigma_t \: \kappa_t \: \mathbf n \right)   }
\end{array}
\right.
\label{source}
\end{eqnarray}

The  system of equation (\ref{discrete}) is written with quantities that are expressed only with the fundamental units of length and time, whereas the dedicated Navier-Stokes, Navier-Lam{\'e} and Maxwell equations involve all the fundamental units, length $L$, time $T$, mass $M$, intensity $I$ and sometimes temperature $\Theta$.

\subsection{Correspondence between physical variables} 

Table (\ref{corres}) presents, for phenomena supposed to be described by the equation (\ref{discrete}), mechanics of fluids and solids and electromagnetism, the correspondence between the variables and the properties conventionally used for each of the domains described and those to be fixed in the equation, $dt \: c_l^2$ and $dt \: c_t^2 $. The quantities in this table are respectively $\chi_T$ the coefficient of isothermal compressibility, $\mu_f$ the viscosity of the fluid, $\nu_f$ its kinematic viscosity, $\lambda$ and $\mu_s$ the Lam{\'e} coefficients,  $\varepsilon_m$ the permittivity, $\mu_m$ the magnetic permeability and $\sigma_m$ the electrical conductivity. The variable of the discrete equation is the velocity $\mathbf V$ of the particle or medium, $\mathbf U$ is the displacement of the solid, $e$ is the electric potential, $\bm j$ is the density of electrical current and charge density is noted $\rho_m$.
\begin{table}[!ht]
\begin{center}
\begin{tabular}{|c|c|c|c|c|c|c|}   \hline
          &   $\mathbf V$  &$\phi$  &   $\bm \psi$      & $dt \: c_l^2 $   &  $dt \: c_t^2 $  \\ \hline  \hline
fluids & $\mathbf V$  & $ p / \rho $ & $\bm \omega / \rho$ & $dt / ( \rho \: \chi_T ) $   & $ \nu_f = \mu_f / \rho $  \\ \hline
solids  & $\mathbf U /dt $  & $ p / \rho $ & $\bm \omega / \rho$ & $dt \: (\lambda + 2 \: \mu_s) / \rho $ & $\nu_s = \mu_s / \rho $ \\ \hline
electro.    & $\bm j / \rho_m$  & $ (\rho_m  /  \rho) \: e$  & $(\rho_m / (\rho \: \sigma \: \mu_m)) \: \mathbf B$ & $dt / ( \varepsilon_m \: \mu_m ) $   & $\nu_m = 1/ (\mu_m \: \sigma_m) $  \\ \hline
\end{tabular}
\caption{ \it Correspondence of quantities, variables and properties used in discrete mechanics and the usual quantities in mechanics and electromagnetism where $\mathbf V$ is velocity, $\mathbf U$ displacement, $p$ pressure, $\bm \omega$ the constraint, $\bm j$ current density, $\rho_m$ electrical charge density and $\mathbf B$ the induction magnetic field. }
\label{corres}
\end{center}
\end{table}

From this table, it is easy to extract (\ref{corres}) all the electromagnetic quantities from the potential ones $\phi$, $\bm \psi$ and the velocity $\mathbf V$. Thus, the curl of the vector potential is equal to $\nabla \times \bm \psi = \nabla \times \left((\rho_m / (\rho \: \sigma \: \mu_m)) \: \mathbf B \right) $ and we find the Maxwell-Thomson law $\nabla \cdot \mathbf B = 0$, i.e. the fact that the magnetic field has no charge. As in mechanics, the physical quantities are integrated within the operator gradient and dual curl and they do not have to be derived in space, they are constant on the whole primal surface $\mathbf S$. It can be seen that the correspondence between mechanical and electromagnetic variables has nothing in common with the analogies presented in the literature. In discrete mechanics, the equation is the same and the potentials of a true Hodge-Helmholtz decomposition are those of a single quantity, the acceleration.

Density $\rho$ is that of the fluid medium $\rho_f$ or the solid $\rho_s$ or the vacuum $\rho_v = 0$; even in the latter case, the potentials continue to make sense. For example, for a perfect gas, $\phi = p / \rho = r \: T$ continues to have a value as long as the notion of temperature continues to make sense. For the electromagnetic phenomena in the vacuum, it is the current $\bm j$ which becomes null at the same time as the density. These cases correspond to very compressible media for which the divergence of velocity is very great; deletions of terms {\it a priori} are to be applied with great care, as the product of two terms, one of which tends to zero and the other to infinity, is of course undetermined. The system (\ref{discrete}) is unsteady and applies in all the cases previously mentioned, whatever the time-lapse $dt$ considered: for values compatible with the physics of phenomena, including the propagation of light, the system will report evolutions over time. For much larger values of $d t$, the evolution will not be physical but the convergence state will correspond to the stationary solution of the problem. Given the highly implicit character of the discrete equation, the formulation is very robust.

In electromagnetism, the velocity in the vacuum $c_0$ is equal to the celerity of light and there is no transverse celerity. There are, however, polarizable transverse gravitational waves whose velocity is currently fixed at longitudinal velocity $c_0$; in the absence of different information, we will use this result. It is recalled that the velocity of the particles (matter, electrons, photons) is not limited and that the celerities of the associated waves are exclusively measured quantities. The quantity $p^o$ is the equilibrium mechanical pressure and $\bm \omega^o$ the perfectly defined shear-rotation stress in fluid and solid.

The attenuation factors $\alpha_l$ and $\alpha_t$, which are dimensionless quantities between $0$ and $1$, are also intrinsic properties of the media. These factors depend largely on the time constants $\tau_f$, $\tau_s$ and $\tau_m$ corresponding to the relaxation times of the transverse phenomena.
For example, for water, if $\nu_f = \mu_f \: \rho \approx 10^{-6}$ is the kinematic viscosity and $c_l^2 = 1 / (\rho \: \chi_T) \approx 2.25 \: 10^6$, the characteristic time is then $\tau_f = \nu_f / c_l^2 \approx 10^{-12} s$. It should be noted that transverse celerity is not known for water. It is understood that water relaxes the shear stresses on characteristic times greater than $\tau_f$, and we can adopt $\alpha_t = 0$. For elapsed-times $dt$ of this order of magnitude, the accumulation of shear stresses is no longer negligible. The same analysis can be made for the accumulation of constraints in a solid where we have $\nu_s \approx 2 \: 10^7$ and $c_l^2 \approx 5 \: 10^6$ that is $\tau_s \approx 0.25$ in copper; in this case, copper accumulates the shear stresses $\alpha_t = 1$. For dielectric materials, the values are highly variable and it is necessary to perform a preliminary analysis.

The transformation of mechanical or electrical energy into heat is due to the viscous friction described by the dual curl but also by the term $- \kappa \: \mathbf V$ of the equation (\ref{discrete}). This dissipation is evaluated by the function $\Phi_d = dt \: c_l^2 \: (\nabla \cdot  \mathbf V)^2 + dt \: c_t^2 \: (\nabla \times \mathbf V)^2 + \kappa \: | \mathbf V |^2$ in discrete mechanics. In electromagnetism, this latter contribution corresponds to Joule's law which is written in this context $- \nabla \phi \cdot \mathbf t = \kappa \: | \mathbf V |^2$ by linking the potential difference to the dissipation. 

Contrary to what might be thought, terms with very large or very small coefficients cannot be eliminated. For example, if we want the flow to be incompressible, we must keep $\nabla \cdot \mathbf V$ in the equation of motion, and in fact it is when the longitudinal velocity is very high that the divergence becomes very low, the term $dt \: c_l^2 \: \nabla \cdot \mathbf V$ is an order of magnitude of the other terms of the equation, {\it a priori} of order one. These factors make it possible to maintain persistent effects in the absence of any velocity, such as the magnetic field of a magnet in the very long term. Its demagnetization and the hysteresis induced by a current will be taken into account naturally by the equation of motion.

The system (\ref{discrete}) and properties (\ref{corres}) are sufficient to deal with any problem in one of the domains mentioned. The variables are the scalar potential $\phi$ defined at the ends of the edges $\Gamma$ and the vector potential $\bm \psi$, a pseudo-vector in continuum mechanics, associated with the normals $\mathbf n$ of each of the primal facets; the number of facets having the $\Gamma$ segment in common is variable and the vector, tensor or quadrivector formulation no longer makes sense in this discrete context. It is of course possible to return to the usual variables, for example electric potential $e$, electric field $\mathbf E$, magnetic field $\mathbf B$, excitation field $\mathbf H$, magnetization $\mathbf M$, charge density $\rho_m$ or current density $\bm j$, etc. for electromagnetism where $c^2 = 1 / \varepsilon \: \mu $. All these quantities which are not independent have been defined over time and have become usual notions, but it is no less legitimate to consider only $\phi$ and $\bm \psi$.

It is necessary to add a law of conservation on a particular potential, density for fluids $\rho_f$ and solids $\rho_s$ and $\rho_m$ density of charge in electromagnetism; the conservation law is found for all cases treated:
\begin{eqnarray}
\displaystyle{  \frac{d \rho }{dt}  + \rho \: \nabla \cdot \mathbf V = 0 }
\label{masse}
\end{eqnarray}

In a discrete approach, for a process of temporal accumulation between two states at the instants $t^o$ and $t^o + dt$, the conservation law is expressed by:
\begin{eqnarray}
\displaystyle{ \rho  = \rho^o - \rho^o \: dt \: \nabla \cdot \mathbf V }
\label{massed}
\end{eqnarray}

The value of the pressure on the points can be deduced from it explicitly, whatever the geometrical topology (polyhedra and polygons with any number of faces); if there is a discontinuity of the density at $c$ (Figure \ref{discreet}) on the edge $\Gamma$, the pressure at $b$ can be deduced from its value at $a$ by the discrete integration:
\begin{eqnarray}
\displaystyle{ p_b = p_a -  \int_a^c \rho_1 \: \nabla \phi \cdot \mathbf t \: dl -  \int_c^b \rho_2 \: \nabla \phi \cdot \mathbf t \: dl }
\label{integrec}
\end{eqnarray}
where $\nabla \phi$ is constant on the edge.

The solution to any problem is to find $\left(\phi, \bm \psi, \rho \right)$ as a function of space and time. These quantities corresponding to each equilibrium, defined as the exact satisfaction of equation (\ref{discrete}), are persistent, and stopping the integration process in time will not modify these values.
Physical properties are also updated if they depend on variables and time.

\section{Numerical methodology}

\subsection{A ready to use formulation} 

The operators of the vector equation of discrete mechanics already have a geometric meaning in a three-dimensio-nal space.
These discrete operators can be defined simply from the basic topology presented in figure (\ref{discreet}). First, the discrete gradient is calculated as a difference, for example the scalar potential gradient $\phi$ will be written $\nabla \phi = (\phi_b - \phi_a) / d$. It can be seen from the outset that the gradient vector is not that of continuum mechanics and represents a scalar oriented in the direction $\mathbf t$. The primal curl of vector $\mathbf V$ is calculated as the circulation over all the edges of the oriented surface $\mathcal S$ with $\nabla_p \times \mathbf V$ and will be carried by the unit vector $\mathbf n$. The divergence of a vector, for example, is expressed at a point from the flows of the different oriented segments that converge towards it. The fourth operator is the dual curl $\nabla_d \times \bm \psi$ where the components of $\bm \psi$ are orthogonal to the primal surfaces $\mathcal S$. 
It should be noted that the 2D/3D distinction does not exist. Indeed, even for a planar primal topology, the vector $\bm \psi$ is carried by the unit vector $\mathbf n$ orthogonal to this surface.

The two operators, gradient and dual curl, are those that project the action of different effects on the $\Gamma$ segment. This oriented edge is also the one on which the conservation of the acceleration will be carried out and where the various vector quantities will be evaluated, in particular the components $\mathbf V$ of the velocity.

In the selected topological structure, some operators are exact in the sense that the numerical error committed to evaluate them in a discrete point of view is zero. This is the case of the gradient which is defined by a difference and of the primal curl which is calculated from the Stokes theorem as the circulation of the vector on the contour $\Gamma$. The two other operators, divergence and dual curl, induce numerical errors that depend on the quality of the mesh used and the way in which the dual space is built.

Whereas classical mechanics has been established mainly by considering the divergence theorem for the relation between a flux on a surface and a volume and then making the elementary control volume tend to zero to obtain a formulation at a vertex, the mechanics of discrete media derive the equation of motion from the fundamental theorem of analysis and its consequences, {\it i.e.} the Stokes theorem in particular.
These fundamental theorems are briefly recalled.
If $\mathbf F(x)$ is uniformly and continuously differentiable over $[a, b]$ the fundamental theorem of the analysis or fundamental theorem of differential and integral calculus is written:
\begin{eqnarray}
\displaystyle{  \mathbf F'(x) = \mathbf f(x) \rightarrow \int_a^b f(t) \: dt = F(b) - F(a)}
\label{theorem1}
\end{eqnarray}

The Stokes theorem which follows from the previous theorem makes it possible to calculate the rotational linked to a surface as the circulation on its contour:
\begin{eqnarray}
\displaystyle{  \int_{\Gamma} \mathbf V \cdot \mathbf t \: dl = \int \!\!\!\!\! \int_{\Sigma} \nabla \times \mathbf V \cdot \mathbf n \: ds  }
\label{theorem1b}
\end{eqnarray}

This theorem is of particular scope in a discrete medium and provides the possibility of calculating the rotational on a surface without knowing explicitly the velocity vector itself. It is enough to know its components on a closed contour. The notion of a referential for defining a velocity vector at a point becomes less essential. Moreover, since the rotational is not defined at a point or on a line, this operator is only defined by passing to the limit in continuum.

The divergence or Green-Ostrogradski theorem derived from the Stokes theorem will make it possible to calculate the upgrade of the quantities defined at a given point, such as scalar potential or density:
\begin{eqnarray}
\displaystyle{  \int \!\!\!\!\! \int_{\Sigma}  \mathbf V \cdot \mathbf n \:  ds =  \int \!\!\!\!\!  \int \!\!\!\!\! \int_{\Omega}  \nabla \cdot \mathbf V  \:  dv  }
\label{theorem2}
\end{eqnarray}

The derivation of the discrete equation of motion is performed on an edge $\Gamma$. As a consequence, it is essential to provide an interpretation of the product of two functions on it. A version of the finite increments theorem is the generalized mean value theorem (consequence of Rolle's theorem) for the integrals: $\exists \: c \in [a, b]$ such that:
\begin{eqnarray*}
\displaystyle{   \int_a^b f(x) \: g(x) \: dx = f(c) \: \int_a^b  g(x) \: dx }
\label{theorem3}
\end{eqnarray*}

This theorem is valid in one dimension. It is particularly suited to the approach developed here. For two-phase flows, the value of the density on a edge can thus be framed from its values at the vertices $a$ and $b$.

Two important properties of continuum, $\nabla \times \nabla \phi = 0 $ and $\nabla \cdot \nabla \times \bm \psi = 0$ are indispensable in discrete mechanics.
If we denote the discrete quantities by the index $h$, it is easy to show, on the primal topology, that the discrete curl of a discrete gradient is zero:
\begin{eqnarray}
  \left\{ 
\begin{array}{llllll}
\displaystyle{   \int_a^b  \nabla \phi  \cdot \mathbf t \: dl = \phi_b - \phi_a } \\  \\
\displaystyle{  \int_{\Gamma}  \nabla \phi  \cdot \mathbf t \: dl  = 0 } \\  \\
\displaystyle{  \int \!\!\!\! \int_{\cal S} \nabla \times \big( \nabla \phi \big) \cdot \mathbf n \: ds  = 0 } \\  \\
\displaystyle{  \nabla_h \times \big( \nabla_h \: \phi \big)  = 0 }
\end{array}
\right.
\hspace{10.mm}
  \left\{ 
\begin{array}{llllll}
\displaystyle{ \sum_{i=1}^n \:  \Gamma_i = \sum_{i=1}^n \:  \int \!\!\!\! \int_{s} \nabla \times \bm \psi \cdot \mathbf n \: ds = 0 } \\  \\
\displaystyle{  \int \!\!\!\! \int_{\cal S} \big( \nabla \times \bm \psi \big) \cdot \mathbf n \: ds  = 0 } \\  \\
\displaystyle{  \int \!\!\!\! \int \!\!\!\! \int_{\cal V} \nabla \cdot \big( \nabla \times \bm \psi \big) \: ds  = 0 } \\  \\
\displaystyle{  \nabla_h \cdot \big( \nabla_h \times \bm \psi \big)  = 0 }
\end{array}
\right.
\label{propriop}
\end{eqnarray}

\begin{figure}[!ht]
\begin{center}
\includegraphics[width=4.2cm,height=3.6cm]{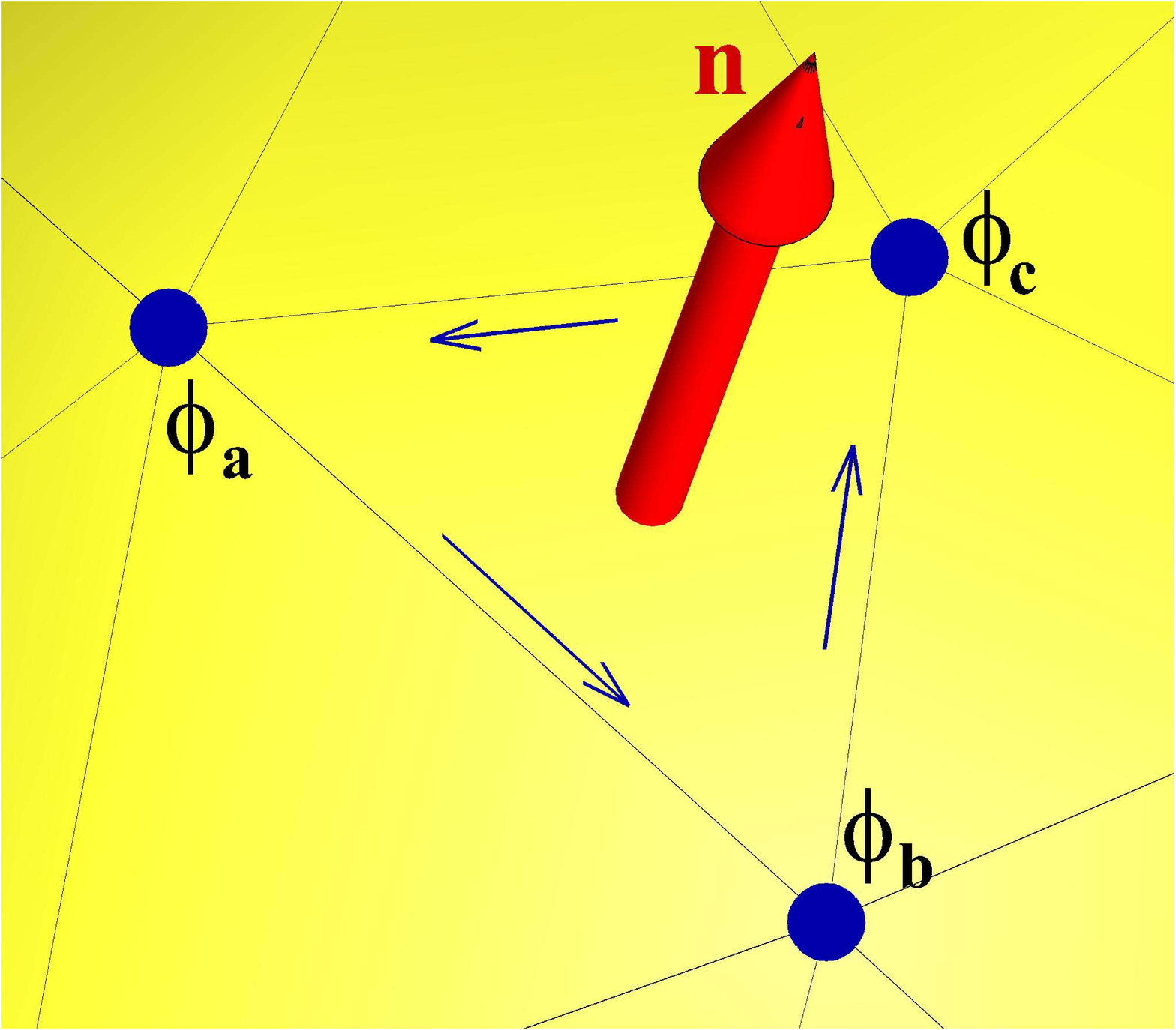}
\hspace{10.mm}
\includegraphics[width=4.cm,height=4.2cm]{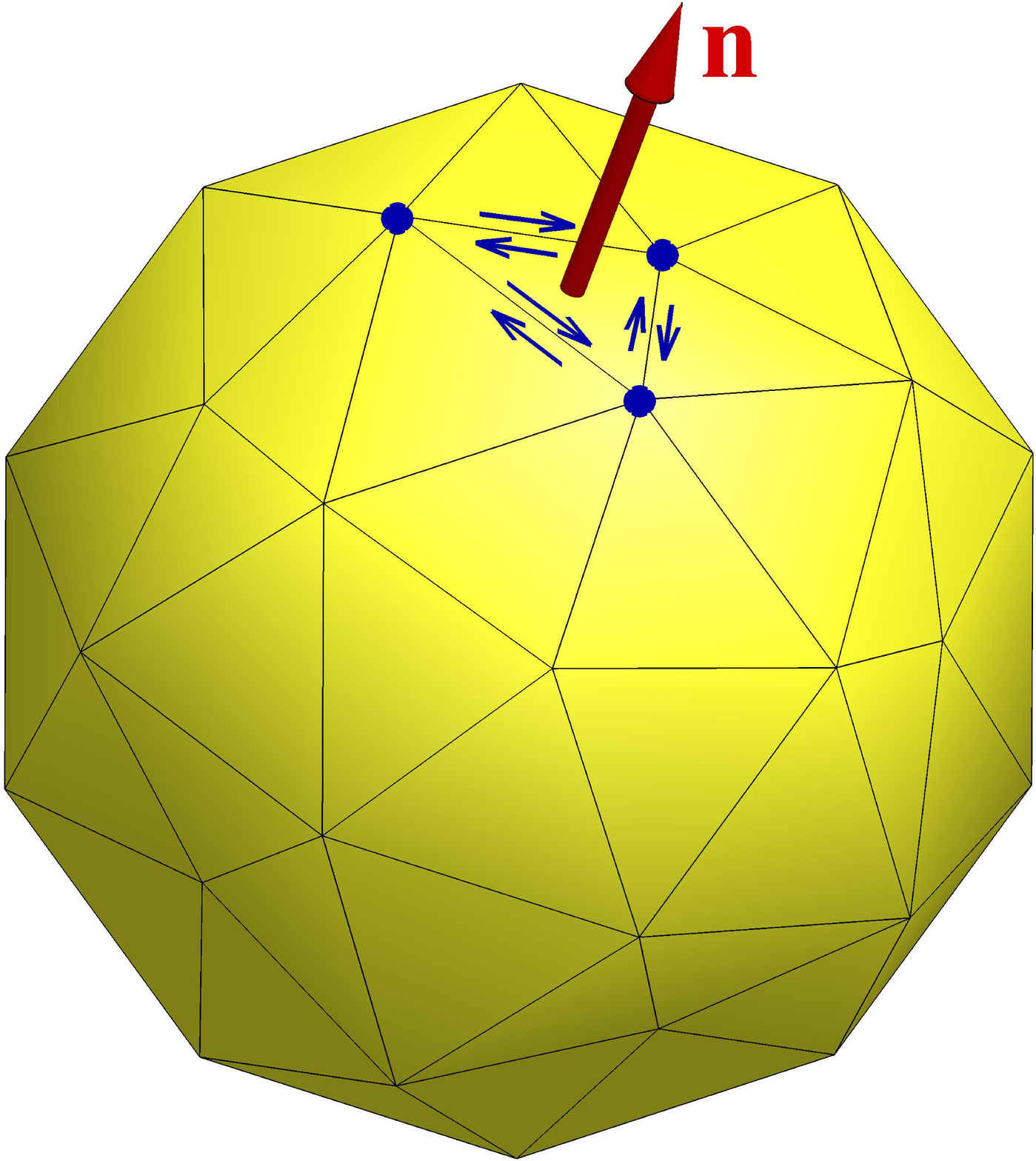}
\vspace{-1.mm}
\caption{\it Operators properties; at left $\nabla \times \nabla \phi= 0$ and at right $\nabla \cdot \nabla \times \bm \psi= 0$ }
\label{propri}
\end{center}
\end{figure}

Similarly, the discrete divergence of the discrete primal curl calculated on the dual volume is zero.
Figure (\ref {propri}) shows how the property $\nabla_h \times \nabla_h \phi = 0$ is checked on the primal topology and how $\nabla_h \cdot \nabla_h \times \bm \psi = 0$ is checked on the dual topology.

All the previous properties enable the equation of discrete motion to formulate acceleration as the sum of a gradient and a rotational, that is, formally, as a Hodge-Helmholtz decomposition.
This decomposition is mainly used for the resolution of the Navier-Stokes equations for which it separates an irrotational contribution from a solenoidal one for any vector. Thus, the correction of the velocity associated with the irrotational part makes it possible to construct a divergence-free field, \cite{Ang12aa}, \cite{Cal15c}. In the general case, managing a decomposition into two orthogonal terms satisfying boundary conditions can be applied to many other fields, such as imaging, fingerprint recognition, and so on. Since any vector can be split into these two terms, it seems natural to look for the components of any vector coming {\it a priori} from a physical field. For example, velocity and acceleration are subject to this general rule.

Some operators that are combinations of the previous essential operators, including the Laplacian $\nabla^2 \phi = \nabla \cdot \nabla \phi$ or $\nabla^2 \bm \psi = \nabla \cdot \nabla \bm \psi$, will not be used. They can induce {\it artifacts} or, for vectors, an increase in the tensorial order of the operators. Tensors of order equal to or greater than two do not have an appropriate representation in the context of discrete media. For example, the gradient of a vector $\nabla \mathbf W$ which has a clear meaning in a continuum cannot be represented in discrete mechanics. Therefore, it will be essential to know if all the physical behaviors can be described by terms associating only the essential operators.

\vspace{3.mm}

It may be necessary to ask the question of the necessity of using tensors to describe all the observed behaviors and all the physical phenomena coming from a macroscopic vision. Even if it seems legitimate to represent the anisotropic behavior of a material with the help of a second-order tensor, the laws of mechanics are not necessarily described by them. The mechanics of solids, general relativity and fluid mechanics, {\it etc} make extensive use of higher-order tensors in the laws they have generated. For example, in solid mechanics, shear stress is defined as the gradient of a vector, whereas it can be simply written in rotational terms. Over time, there has been a confusion between the constitutive laws, state laws and fundamental laws supposed to describe physics. The equations of physics can be written indifferently in vectorial or tensorial form, the typical example being that of Maxwell's equations; each formulation has its advantages and disadvantages.

Spatial convergence depends on the operators: the gradient and primal curl are exact ,whereas the two other operators are of order two and spatial convergence is therefore of order two whatever the polygonal or polyhedral elements that are chosen. In all cases where there is an analytical solution represented by a polynomial of second degree, then the numerical solution is exact to almost machine precision. The second-order convergence has been found in many cases of numerical simulations using unstructured meshes. Temporal convergence is illustrated by the validation case below.

The algebraic formulation is directly derived from the equation of discrete motion (\ref {discrete}); the unknowns of the linear system are the scalar values $V = \mathbf V \cdot \mathbf t$ of each oriented edge  $\Gamma$ composing the primitive geometric topology. The term of inertia in $ | \mathbf V |^2/2$ is linearized but all other terms in the equation are fully implied. The resolution of the linear system is ensured by means of a non-preconditioned conjugate gradient solver BiCGStab2. The solution on $\mathbf V$ allows the calculation {\it a posteriori} of the divergence $\nabla \cdot \mathbf V$ and the primal curl $\nabla \times \mathbf V$, and the potentials $\phi$ and $\bm \psi$ and possibly the displacement $\mathbf U$ are then explicitly upgraded.

\subsection{The Green-Taylor vortex} 

The Green-Taylor vortex is a synthetic analytical solution of the Navier-Stokes equations that corresponds to a case of incompressible unsteady flow. This case is often used to compute convergence orders in time and space of numerical methods. It is considered here to show that the discrete formulation makes it possible to find the second-order analytical solution in time and space. More particularly, it should enable us to understand the role played by the different terms of the discrete motion equation and also to characterize how they combine to satisfy operator properties identically.

The equation of the unsteady movement reads:
\begin{eqnarray}
\hspace{-10.mm}
\displaystyle{   \frac{\partial \mathbf V}{\partial t} + \nabla \phi_i - \nabla \times \bm \psi_i   =  - \nabla \left(  \phi^o  - r \: \nabla \cdot \mathbf V \right) - \nabla \times \left(  \nu \: \nabla \times \mathbf V  \right) + \mathbf S_{\mu} + \mathbf S_i }
\label{green}
\end{eqnarray}

where $\phi_i = | \mathbf V | ^2/2$ and $\bm \psi_i = \phi_i \: \mathbf n$; the quantities $\mathbf S_i = \partial \mathbf V / \partial t$ and $\mathbf S _ {\mu}$ are suitable source terms that lead to a stationary solution. Parameter $r$ makes it possible to maintain, at any instant, the divergence of the velocity below $1 / r$; this can be considered as zero throughout the simulation. Since the medium is a Newtonian fluid, the accumulation of shear-rotation constraints is zero $\bm \psi^o = 0$ with the instantaneous potential being equal to $\bm \psi = - \nu \: \nabla \times \mathbf V$. In fact, the dual curl of the vector potential is equal to source term $\mathbf S_{\mu}$ for this form of the equation.

For $x \in [-0.5, 0.5]$, the solution is written:
\begin{eqnarray}
\hspace{-5.mm}
\left\{  
\begin{array}{llllll}
\displaystyle{ \phi =  \frac{V_0^2}{2} \: \left( cos(\pi x)^2 + cos(\pi y)^2 \right) \: \left( 1 - \exp(-\pi \: t) \right)^2 } \\ \\
\displaystyle{ u = - V_0 \: cos(\pi\: x) \:\: sin(\pi \:y)  \: \left( 1 - exp\left( - \pi \: V_0 \: t \right) \right)  }  \\  \\
\displaystyle{ v = \:\:\: V_0 \: sin(\pi\: x) \:\: cos(\pi \:y) \: \left( 1 - exp\left( - \pi \: V_0 \: t \right) \right)  }
\end{array}
\right.
\label{taylor}
\end{eqnarray}
with $\mathbf V = u(x, y) \: \mathbf e_x + v(x, y) \: \mathbf e_y$ and $\bm \psi = - \nu \: \nabla \times \mathbf V$.
 
The source term calculated from solution (\ref{taylor}) leads to a separation between inertial and viscous terms:
\begin{eqnarray}
\left\{  
\begin{array}{llllll}
\displaystyle{   \frac{\partial \mathbf V}{\partial t} = \mathbf S_i }  \\  \\
\displaystyle{   \nabla \times \left(\bm \psi^o - \nu \: \nabla \times \mathbf V \right)  = 0 }  \\  \\
\displaystyle{ - \nabla \times  \bm \psi_i  = - \nabla \phi_B  }
\end{array}
\right.
\label{green2}
\end{eqnarray}

The potential of Bernoulli $\phi_B = \phi + \phi_i$ is the equivalent of the Bernoulli pressure. We find that $\nabla \times \bm \psi^o = \mathbf S_{\mu}$. Adding this source term into the motion equation amounts to imposing a vector potential $\bm \psi^o$ depending on time. In discrete mechanics, equilibrium is not required by component and only the addition of accelerations on the edge $\Gamma$ makes it possible to translate the mechanical equilibrium. Here, the solution of the problem $(\mathbf V, \phi, \bm \psi) $ is obtained directly by the resolution of the vector equation (\ref{green}).

The flow of Green-Taylor is often used to obtain the order of convergence in time and space of a given numerical methodology. Numerous reference cases show that the discrete formulation provides a second-order precision in space and time. Table (\ref{green-temps}) gives only the evolution of the numerical error in time according to the number of elapsed-time $dt$.
\begin{table}[!ht]
\small
\begin{center}
\begin{tabular}{|c|c|c|c|c|}   \hline
   Time step &    Error on $\mathbf V$        &   Error on $\phi$          \\ \hline  \hline
  $10^{-2}$         & $7.166 \: 10^{-3}$ & $1.194 \: 10^{-2}$  \\ \hline
  $4 \: 10^{-3}$    & $1.667 \: 10^{-3}$ & $2.153 \: 10^{-3}$  \\ \hline
  $10^{-3}$         & $1.109 \: 10^{-4}$ & $1.152 \: 10^{-4}$  \\ \hline
  $4 \: 10^{-4}$    & $1.723 \: 10^{-5}$ & $2.467 \: 10^{-5}$  \\ \hline
  $10^{-4}$         & $9.902 \: 10^{-7}$ & $1.367 \: 10^{-6}$  \\ \hline
  $4 \: 10^{-5}$    & $1.485 \: 10^{-7}$ & $2.246 \: 10^{-7}$  \\ \hline
  $ 10^{-5}$        & $1.060 \: 10^{-8}$ & $1.684 \: 10^{-8}$  \\ \hline
  Order             &            $2$                 & $2$                         \\ \hline
\end{tabular}
\caption[Green-Taylor Vortex]{\it Green-Taylor vortex; error in time on $L_2$ norm, the time reached is equal to $t = 10^{-2}$ and the mesh of $64^2$ cells to $1024^2$ cells to ensure the non-saturation of the error in time by the error in space. }
\label{green-temps}
\end{center}
\end{table}

Convergence in time at order two is obtained using a Gear scheme for the unsteady term, the linearization of the terms of inertia $| \mathbf V |^2/2$ in the form $(\mathbf V^{n} \cdot \mathbf V^{n + 1}) / 2$ and the potential upgrade by the expression $\phi^{n + 1 } = \phi^{n-1} - 2 \: \nabla \cdot \mathbf V^{n + 1}$.

The use of the vector of Lamb $\mathbfcal {L} = \nabla \times \mathbf V  \times \mathbf V$ in continuum mechanics leads to the expression of the Navier-Stokes equation by component. However, this vector makes it possible to represent the inertial term in a planar surface $(x, y)$ for a 2D description.
It should be noted that the divergence of the Lamb vector $\mathbfcal{L}$ is not zero, even if its rotational $\nabla \times \mathbfcal {L} = 0$ is really null, so that the Lamb vector derives from a scalar potential of the considered problem. In three dimensions of space, the Lamb vector is more difficult to interpret. This leads to different turbulence properties in 2D and 3D. Like all simulations performed in incompressible motions, the results of the discrete model are identical to those of the Navier-Stokes equation.

\section{Applications} 

The equation system of Discrete Mechanics (DM) (\ref{discrete}) is representative, without modification, of several domains of physics, fluid mechanics, solid mechanics, wave propagation and heat transfer, to cite just a few.

Numerous examples have shown the validity of the system (\ref{discrete}) in fluid mechanics. For example, compressible, incompressible, non-isothermal or two-phase flows can be approximated with DM to recover the standard known results, in particular the classical analytical solutions of Poiseuille or Couette flows.  Reference cases, such as the lid-driven cavity, the backward-facing step or the flow around a cylinder, make it possible to show that the DM model converges to order two in space and time for both velocity and pressure. The flows associated with heat and mass transfers including multi-components are reproduced in a similar way. More complex problems of shock waves, like the Sod tube, phase changes, boiling and condensation \cite{Ami14} are treated in a coherent way by integrating discontinuities within the equations of motion. Two-phase flows with capillary effects, surface tension or partial wetting, are particularly well suited to the discrete mechanics model \cite{Cal15c}.

In the present form, the system (\ref{discrete}) is relatively close to the Navier-Lam{\'e} equation associated with the study of stresses and displacements in solids. It differs, however, on several points: more particularly, the discrete formulation is established in velocity and the displacement is only an accumulation of $\mathbf V \: dt$, as the velocity is itself the elevation of $\bm \gamma \: dt $. Numerous examples of simple solicitations make it possible to find the solutions of the Navier-Lam\'e equation. More complex 2D and 3D problems on monolithic fluid-structure interactions have already made it possible to validate the proposed formulation \cite{Bor14}, \cite{Bor16}. The vision of a continuous memory medium makes it possible to treat the problems of large deformations and large displacements in a formulation where pressure stress and shear are obtained synchronously without compatibility conditions. Given the original dissociation between compression effects and rotation, the material frame-indifference introduced by Truesdell \cite{Tru74} is satisfied naturally. The complex constitutive laws can be treated without difficulty and only the physical parameters written in the equation of motion must be known.

\subsection{Fluid Mechanics} 

\subsubsection{Lid-driven cavity}

The case of the lid-driven cavity, considered at a sufficiently large Reynolds number, is ideal for testing the legitimacy of the inertial term formulation in the equation of motion. From a general point of view, the solenoidal and irrotational parts of inertia are not easily expressed. It seems advisable to present tangible results to the numericians and CFD specialists who are attached to the inevitable Navier-Stokes equations that are widely used. They have shown their relevance in the immensity and variety of the cases modeled and simulated with them over centuries. 
Despite the different physical models and the changes brought by the equation of discrete mechanics that replaces the Navier-Stokes formulation, {\it i.e.} the treatment of pressure that is transformed into Bernoulli pressure, and the writing of the inertia term in the form $ - \nabla \times \left (\phi_i \: \mathbf n \right) + \nabla \left (\phi_i \right) $, the solutions obtained with both approaches are very close to the reference results obtained previously in the literature for the chosen Reynolds number.
\begin{table}[!ht]
\begin{center}
\begin{tabular}{|c|c|c|c|c|c|c|c|c|}   \hline
 Ref.  & $\psi_{max}$ &  $x_{max}$  & $y_{max}$  & $\psi_{min}$ & $x_{min}$  &  $y_{min}$  \\ \hline  \hline
Present $256^2$  & $0.1219$  & $0.5153$  & $0.5352$  &  $-3.086 \: 10^{-3}$ & $0.8040$  &  $0.07310$  \\ \hline
Bruneau al. $2048^2$& $0.12197$ & $0.515465$  & $0.53516$  &  $-3.0706 \: 10^{-3}$ & $0.80566$  &  $0.073242$  \\ \hline
\end{tabular}
\vspace{-1.mm}
\caption{\it Comparison of the results obtained in mechanics of continuous media \cite{Bru06} and those resulting from the DM formulation presented at $Re =  5000$ for a Chebyshev mesh with $256^2$ cells. }
\label{cavity-5000t2}
\end{center}
\end{table}

The results of \cite{Bru06} for a Reynolds number of $Re = 5000$, for which the flow is stationary, are reproduced in table (\ref{cavity-5000t2}). The quantitative comparison concerns the amplitude and position of the vortices generated by the detachment of the flow on the walls of the cavity. Very good accuracy is obtained by comparing the DM formulation and the Navier-Stokes equations to reference \cite{Bru06}. Both formulations provide the same physical solution. 
A convergence study was investigated for this configuration. It showed a spatial convergence rate of $ 2 $ for velocity and pressure. The Bernoulli pressure was used to conduct the evolutions in time and the pressure itself was then extracted from it.
Discrete Mechanics does not bring into question the results obtained with the Navier-Stokes equation. As previously observed, the same solutions are obtained at least with a convergence order and an accuracy that are almost identical to computer errors. However, the most amazing thing is that continuous and discrete models differ on many points. One of these fundamental differences is the use of mass conservation that is always associated with the Navier-Stokes equation. On the contrary, the discrete mechanics equation does not use it explicitly and the mass is always strictly conserved for compressible or incompressible movements: the DM formulation behaves as an autonomous equation that does not require any additional constitutive law. In fact, mass conservation is implicitly integrated into the equation of motion through the term $dt \: c_l^2 \: \nabla \cdot \mathbf V$ \cite{Cal11}.
\begin{figure}[!ht]
\begin{center}
\includegraphics[width=4.5cm,height=4.5cm]{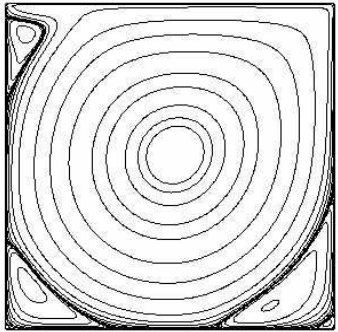}
\caption{ \it Lid driven cavity at $Re = 5000$ with Cartesian mesh (mesh-cart-7.typ2); $\psi_{max} = 0.1211$ at $ x_{max} = 0.5152$ and $ y_{max} = 0.5353$ and $\psi_{min} = - 0.003138$ at $ x_{min} = 0.8024$ and $ y_{min} = 0.07180$.  }
\label{cavite-5000}
\end{center}
\end{figure}

The stream lines shown on figure (\ref{cavite-5000}) are obtained directly from the vector potential $\bm \psi = - \nu \: \nabla \times \mathbf V$ projected on the surface, $\psi = \bm \psi \cdot \mathbf n$.

The lid driven cavity example is a classical flow that shows all the interest of the discrete formulation compared to a continuum type approach. First of all, the solutions with constant properties are strictly the same, whatever the problem dealt with, from the analytical solutions of the Navier-Stokes equations to complex flows whose solutions are obtained numerically. The interest resides in the physical understanding that can be exhibited from the scalar and vector potentials that make the equation of discrete mechanics a true extractor of Hodge-Helmholtz components of acceleration.

\subsubsection{Laplace problem of drop equilibrium}

One of the emblematic cases driven by capillary effects is the equilibrium of a cylindrical or spherical droplet. The problem here is related to the ability of the described methodology to maintain a long-term at-rest state. To obtain this, the Hodge-Helmholtz decomposition leads us to define the surface tension per unit mass $\sigma$ and the curvature $\kappa$ as constants. If the surface tension is not constant, Marangoni type currents develop at the surface of the drop. When the curvature is not constant, orthogonal currents are generated at the interface.

Let us consider the case of a constant mass surface tension $\sigma = \gamma / \rho = 1$. Surface markers are initially seeded on a circle and two cases may occur: on the one hand, the curvature is constant to almost machine precision, which is the case if the markers are arranged exactly on the radius circle $R$, and on the other hand, errors exist on the curvature and in this case, movements of the markers are initiated by the differences in curvature. In the latter case, the presented methodology leads to a repositioning of the markers on the circle.
In the equation of motion, the capillary term is represented by two contributions $\bm \gamma_c = \nabla \phi_c - \nabla \times \phi_c \: \mathbf n$. In the particular configuration where a planar domain is considered, the curvature in the normal direction is zero and the capillary acceleration becomes $\bm \gamma_c = \nabla (\sigma \: \kappa \: \xi)$, where $\xi$ is a phase function related to the $\Sigma$ contact discontinuity of the figure (\ref{discreet}).
For a curvature calculated exactly using a chain of markers, a capillary source term is injected into the equation of motion. As the curvature is constant and not zero, the drop collapses on itself. If the medium is supposed to be incompressible, an equal and opposite force is exerted on the interface due to the pressure increasing inside the interior fluid of the drop. The equilibrium is obtained almost instantaneously and the pressure is uniform inside the drop and equal to $p_c = \gamma / R$ in 2D, $p_c = 2 \: \gamma / R$ for a sphere or $p_c = \gamma \: \kappa$ in general.
\begin{figure}[!ht]
\begin{center}
\includegraphics[width=4.cm,height=4.cm]{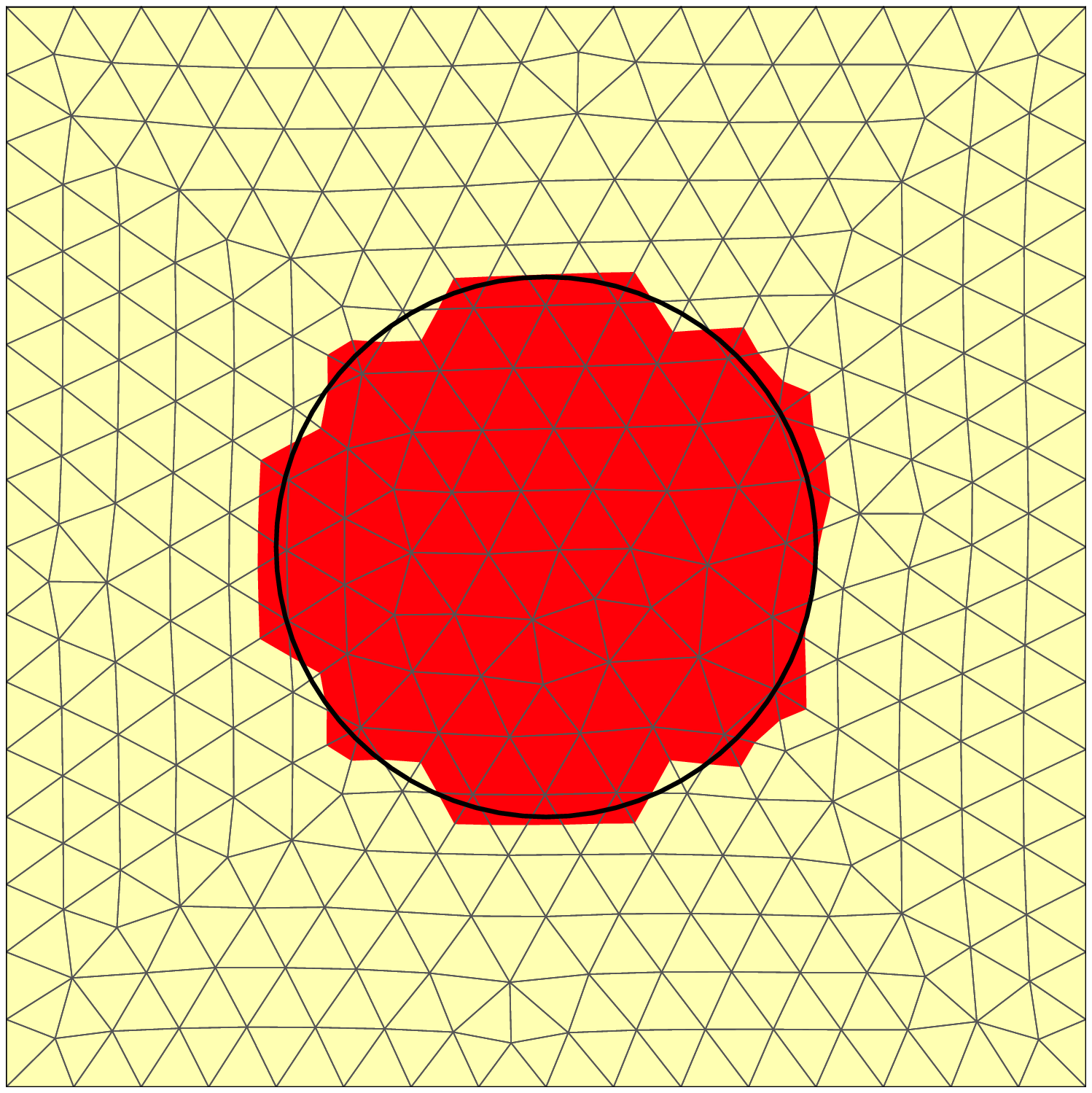}
\includegraphics[width=4.cm,height=4.cm]{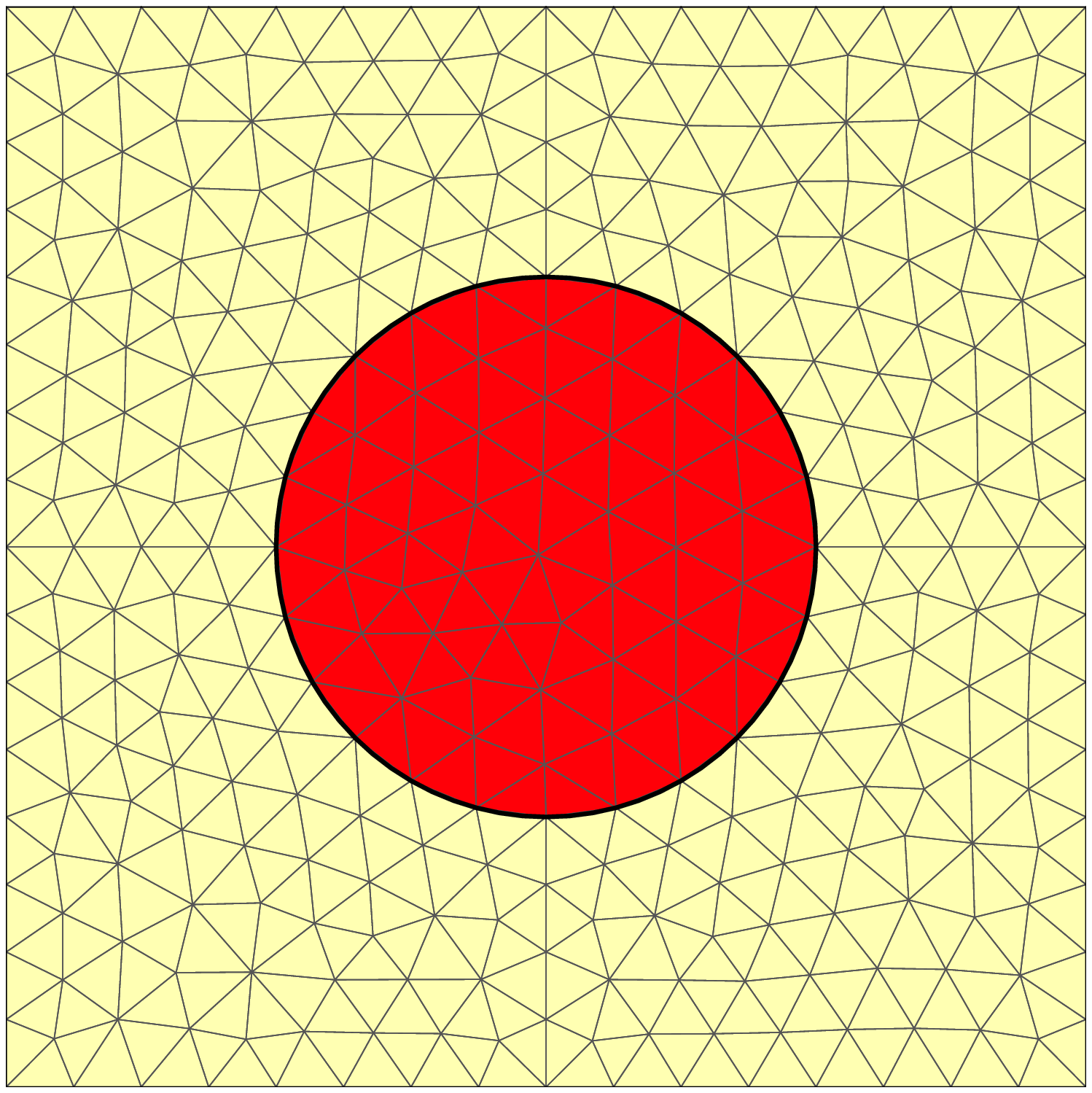}
\caption{\it Capillary potential (or pressure) in a drop of radius $R = 2.5 \: 10^{-3}\: m$ in a square of length $L = 10^{-2}\: m$ for a constant surface tension $\sigma = \gamma / \rho = 1$. On the left, a mesh non-conforming to the disk and a chain of markers to represent the disk are implemented while on the right, an unstructured mesh conforming to the disk is used. In both cases, the pressure is exact and the velocity is strictly zero.}
\label{disque}
\end{center}
\end{figure}

Figure (\ref{disque}) shows two examples corresponding respectively to a structured mesh that does not conform to the circle and another one that conforms to the circular geometry. In both cases the curvature is exactly equal to $\kappa = 1 / R$ and the pressure obtained after one iteration is equal to $p_c = 400 \: Pa$ (for $\rho = 1 \: kg \: m^{-3}$) as expected, while the velocity is zero to almost machine precision both outside and inside the drop.
The same result is obtained with a sphere described by a surface mesh made of triangle elements contained in a cube meshed by regular hexahedrons. It should be noted that this problem is, in the proposed discrete formulation, strictly independent of the density of the two fluids.

The solution of Laplace's problem for a drop is free of any of the parasitic currents widely described in the literature \cite{Sca99}. These instabilities are {\it artifacts} of the physical modeling of the capillary effects \cite{Bra92} in a continuous medium. The discrete approach consists in writing each of the source terms as a Hodge-Helmholtz decomposition, which inevitably leads to feeding one of the two scalar and vector potentials.

\subsubsection{Spreading of a drop} 

The wetting of real surfaces poses many problems of definition, characterization and modeling of the capillary effects in the presence of three media, generally gas, liquid and solid, which join on a contour called the triple line. The flows in the presence of capillary effects involve two characteristics, the surface tension $\sigma_{ij}$ between two media indexed $i$ and $j$ and the local curvature of the interface $\Sigma$ between these supposed immiscible media. The question of the need for an additional physical parameter to introduce the concept of wetting is a legitimate one.
The answer given for many decades was based on observation of the $\theta$ contact angle between two of the media, generally the one formed by the liquid surface and the solid substrate. This is a static measurement obtained from different principles and with various experimental methodologies. When the interface is moving, these approaches lead to the introduction of a contact angle which depends on time, called the dynamic contact angle. This avatar is the object of many laws based on observations of the evolution of the $\theta$ angle over time for specific situations.
Countless references on this subject can be found in the literature. However, as the point of view developed in the present work differs substantially from the various classical approaches, emphasis will be placed on the presentation of the discrete point of view.

We will tackle the problem from the start: as what is sought is the acceleration $\bm \gamma$ of the fluid on the edge $\Gamma$, the actions that affect it can only be written as a sum of contributions. First of all, we can mention the inertial and viscous accelerations, as well as the gravitational acceleration. Capillary acceleration $\gamma_c$ can also be added. The movement of the triple line must therefore be related to all these effects; to define a dynamic contact angle {\it a priori} which depends only on time therefore has no meaning. The modeling proposed in discrete mechanics is based on the following observation: if the $\Sigma$ interface is allowed to evolve freely on the substrate, it will naturally look for its equilibrium position fixed by the static contact angle $\theta$. The dynamics in itself will be handled by the acceleration $\bm \gamma$. The physical behavior of the interface in the zone of the triple line is determined by the need to establish a mechanical equilibrium based on the accelerations of the different pairs of mediums. It is actually the curvature of the interface in this area that best reflects this tendency to return to equilibrium. A specific parameter $\kappa_c$ is introduced, named contact curvature. Once it is known, the instantaneous curvature $\kappa$ will take values that depend on all the effects, while tending to $\kappa_c$ at mechanical equilibrium. For example, a drop of liquid on a solid non-deformable substrate will tend, in the absence of gravity, towards the curvature $\kappa_c$ corresponding to a given static contact angle $\theta$. In this case, this is the entire interface that will have the curvature $\kappa_c$. From the geometry of the interface, it will be possible to link the contact curvature $\kappa_c$ to the contact angle $\theta$. The main advantage of the proposed modeling lies in the fact that no additional parameter is added to the formulation, the curvature is already part of the modeling of $\bm \gamma_c $. On the triple line, it will be enough to impose that the curvature should be equal to $\kappa_c$, the dynamics will be ensured naturally by the equation of motion itself.

The evolution of a water drop on a partially wetting planar surface is a classic problem which has the advantage of corresponding, at equilibrium, to a state at rest.
The initial shape of the droplet corresponds to a semi-circle of radius $R = 10^{-2} \: m$ placed on a horizontal surface and gravity is not taken into account. The surface tension is constant and equal to $\gamma = 1$. Initially, the pressure in the drop is uniform and equal to $p_c = 100 \: Pa$. The motion is generated by imposing a contact curvature $\kappa_c = 24.0143 \: m^{-1}$ on the two vertices of the domain representing the contact line. This value corresponds for the considered geometry to a contact angle of $\theta = 30$ degrees.

\begin{figure}[!ht]
\begin{center}
\includegraphics[width=9.cm,height=2.5cm]{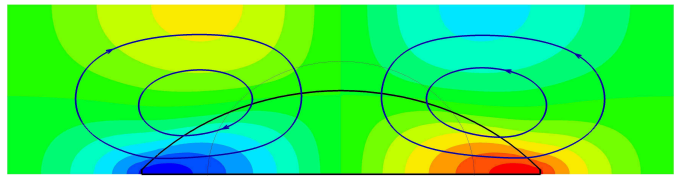} 
\vspace{-3.mm}
\caption{\it Spreading a drop on a slippery surface from the state corresponding to the semicircle; the fields represent the intensity of the horizontal velocity and the stream lines show the circulation of the fluid in the cavity.  }
\label{gouplan-1000}
\end{center}
\end{figure}

The curvature gradient in the vicinity of the triple line causes a movement of the fluid outside and inside the drop which tends to spread it and, as the condition of incompressibility is imposed, the resulting drop volume remains constant and the drop height decreases. Figure (\ref{gouplan-1000}) gives a snapshot of the shape and motion of the fluid. At this instant, the connection condition is not respected either. An instantaneous connection cannot be satisfied and the contact angle cannot be imposed. The contact angle thus varies during the spreading of the drop according to inertial forces, viscosity and capillary forces. The intermediate state is the one shown in figure (\ref{gouplan-1000}).
The convergence towards equilibrium state is rather long. However, when it is reached, the shape of the interface is a portion of circle whose curvature is equal as expected to $\kappa_c$, the velocity is null and the value of the final pressure is equal to $p_c = 24.0143 \: Pa$. 

The treated case corresponds dynamically to a slip surface where $\mathbf V \cdot \mathbf n = 0$. It is of course possible to impose the adhesion of the fluid on the wall $\mathbf V = 0$. Nevertheless, at the considered scales, as the viscosity of the water is low, it does not change the dynamics of the spreading much. It would not be the same on a smaller scale when the dynamic boundary layer is of the order of magnitude of the dimensions of the drop. On the other hand, the progression of the triple line cannot be subjected to the dynamic stress, as it slides freely on the wall according to the forces exerted on it.

It should be noted that in the absence of external forces, for example gravity, the contact angle and contact curvature are linked by simple geometric relations. In this case, it is the surface of the drop $S$ which is initially fixed.
Let us consider the initial surface of the circular cap which is equal to $S = \pi \: R^2/2$. This surface remains constant throughout the simulation. The methodology, in incompressible flow configurations, keeps the volume and mass constant. The effect of the capillary forces on the triple line modifies the local curvature. However, at equilibrium, the shape of the drop corresponds to a circular cap whose characteristics can be calculated easily. If $r$ is the radius of the cap and $\theta$ the contact angle, we obtain:
\begin{eqnarray}
\displaystyle{  S = \frac{r}{2} \: \left( 2 \: \theta - \sin ( 2 \: \theta ) \right) = \frac{1}{2 \: \kappa} \: \left( 2 \: \theta - \sin ( 2 \: \theta ) \right) }  
\label{calotte-a}
\end{eqnarray}

Since $S$ is constant, the contact angle can be calculated using a Newton algorithm by solving equation (\ref{calotte-a}).
For the particular case when $\theta = 0$, the curvature is zero, whereas when the contact angle is equal to $\theta = 180$\textsuperscript{o} and the curvature is equal to $\kappa = \sqrt {2} / R$.
\begin{table}[!ht]
\begin{center}
\begin{tabular}{|c|c|c|c|c|}   \hline
  Curvature          &   Radius of meniscus    &   Contact angle   &  Capillary pressure    \\ \hline  \hline
  $1600 \: m^{-1}$  & $0.625 \: 10^{-3} \: m$  &     $90$\textsuperscript{o}          &   $112 \: Pa$           \\ \hline
  $750 \: m^{-1}$   & $1.333 \: 10^{-3} \: m$  &     $48.2$\textsuperscript{o}        &   $52.5 \: Pa$           \\ \hline
  $2230 \: m^{-1}$  & $0.415 \: 10^{-3} \: m$  &     $150$\textsuperscript{o}         &   $155.04 \: Pa$           \\ \hline
\end{tabular}
\vspace{-2.mm}
\caption{\it Evolution of a drop placed on a planar surface; characteristics of the meniscus at each stage of the simulation.   }
\label{calotte-c}
\end{center}
\end{table}

In this problem, the source term derives from a single scalar potential, capillary pressure. This would no longer be the case if gravity were taken into account: although gravity derives from a potential, the equilibrium surface is no longer a circle and the term $\nabla (\sigma \: \kappa \: \xi) + \mathbf g$ has a non-zero irrotational component. The static equilibrium respects the condition imposed at the contact curvature $\kappa_c$ but the shape of the interface results from the equilibrium of the forces in presence, which only the equation (\ref{discrete}) can restore.
\begin{figure}[!ht]
\begin{center}
\includegraphics[width=5.cm,height=5.cm]{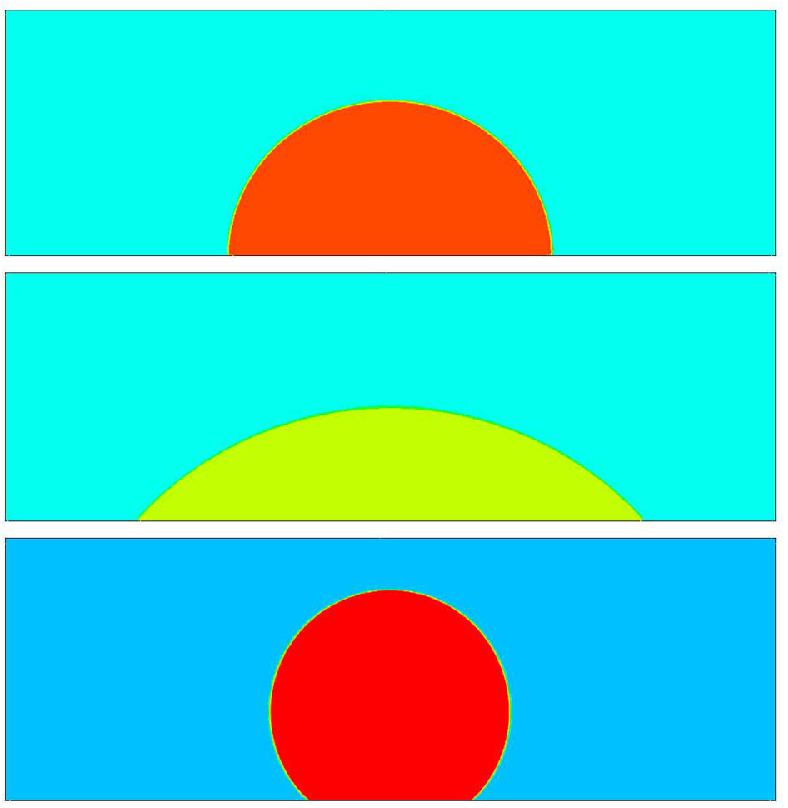}
\hspace{4.mm}
\includegraphics[width=7.cm,height=5.cm]{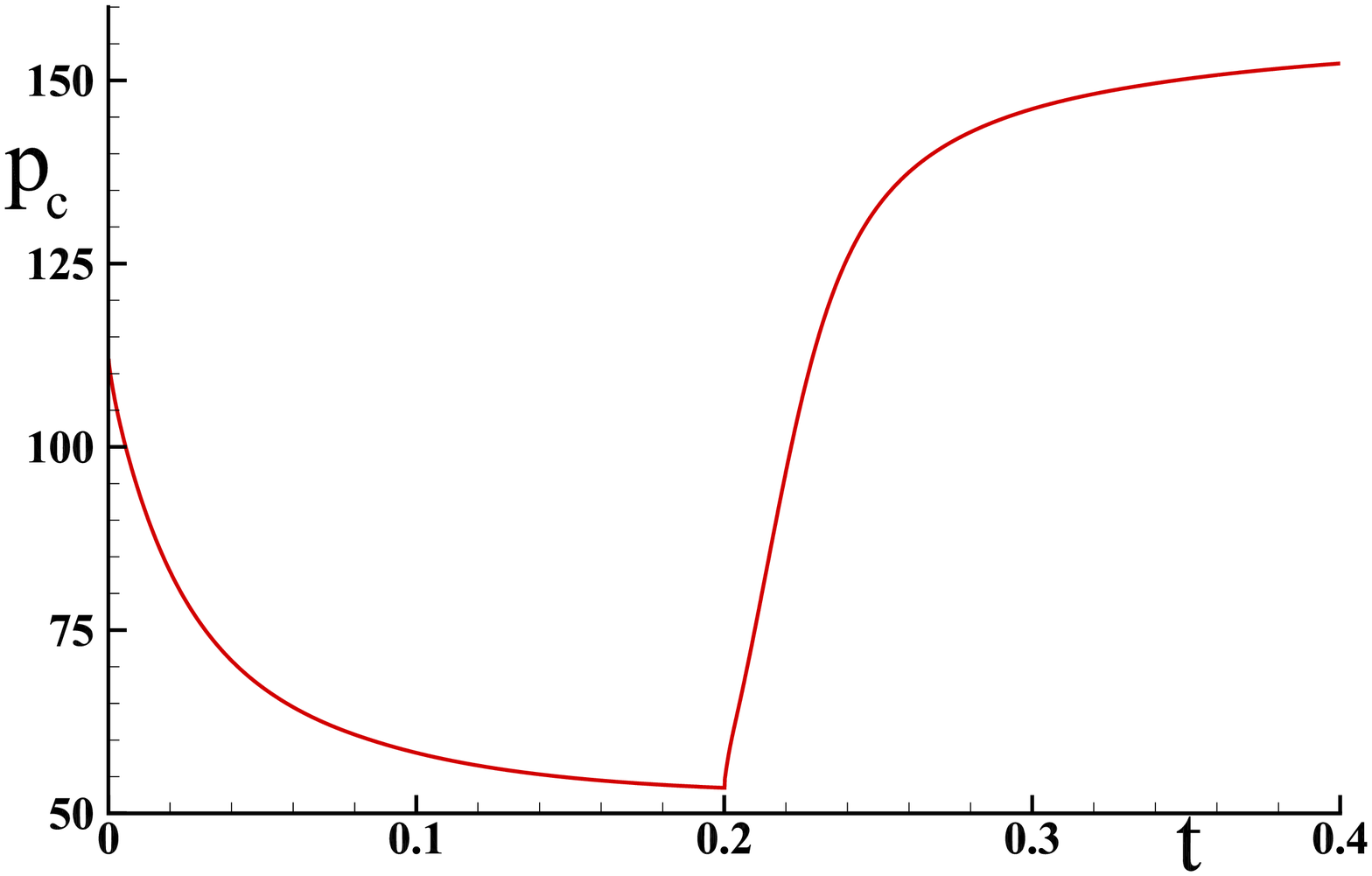} \\
\hspace{-1.mm} (a) \hspace{67.mm} (b) 
\vspace{-1.mm}
\caption{\it Simulation of the evolution of a droplet on a flat surface according to the application of curvature to the wall. (a) At the top, a droplet in equilibrium, where the curvature $\kappa = 1600 \: m^{-1}$ and contact angle $\theta = 90$\textsuperscript{o} are initial conditions, in the center, the curvature $\kappa_c = 750 \: m^{-1}$ ($\theta = 48.2$\textsuperscript{o}) is imposed at the triple line and at time $t = 0.2 \: s$ we impose $\kappa_c = 2230 \: m^{-1}$ ($\theta = 150 $\textsuperscript{o}). (b) Evolution of the average pressure difference  $p_c$ between the drop and the outside medium as a function of time according to variations in the curvature at the wall. The asymptotes of the capillary pressure are at $p_c = 52.5 \: Pa$ and $p_c = 155 \: Pa$. }
\label{calotte-evol}
\end{center}
\end{figure}

The same dynamic problem as in the previous section is now treated with the Arbitrary Lagrangian Eulerian (ALE) method in 2D with an adaptive mesh based on regular triangles. A drop having the shape of a semicircle of radius $R = 0.625: 10^{-3} \: m$ is placed on a planar surface whose wettability can be modified over time from the contact curvature $\kappa_c$. The density of the liquid is chosen equal to $\rho = 1000 \: kg \: m^{-3}$ and its viscosity is $\mu = 10^{-2} \: kg \: m^{-1} \: s^{-1}$, the outside medium being air.
The curvature of the initial semicircle is $\kappa =  1600 \: m^{-1}$. Initially, we impose a contact curvature $\kappa_c = 750 \: m^{-1}$ to simulate a wetting surface and, at a given time $t = 0.2 \: s$ we modify the curvature value which is brought to $\kappa_c = 2230 \: m^{-1}$ to represent a non-wetting surface.
The table (\ref{calotte-c}) gives the characteristics of the droplet for each step of the simulation.

Figure (\ref{calotte-evol}(a)) shows the shape of the drop at each almost stationary state where the velocities are negligible. The pressure inside the drop is then almost constant. As soon as the $\kappa_c$ is modified, the forces exerted on the triple line act to bring the drop back to a state of equilibrium where the contact curvature $\kappa_c$ is satisfied.
Over time, the velocities evolve in the field first to spread the drop to a contact angle of $48.2$\textsuperscript{o} where the velocity fields are similar to those of figure (\ref{gouplan-1000}) and then to contract the interface to a contact angle of $150$\textsuperscript{o}.
Figure (\ref{calotte-evol}(b)) shows the evolution of the pressure difference between the inside and the outside of the drop.
In fac,t the term $\nabla (\sigma \: \kappa \: \xi)$ is an acceleration. It is important to note that at each modification of the contact curvature, the curvature acceleration term has to become zero to reach a state close to equilibrium.

\subsection{Fluid Structure Interaction}

\subsubsection{A simple case}

We consider one of the simplest cases of fluid-structure interaction to study the behavior of two media, one viscous and the other elastic. This test case has a very simple analytical solution that highlights the behavior of the two media modeled with the discrete description (\ref{discrete}). The domain height $h = 1 \: m$ is separated by a $\Sigma$ interface located at $h / 2$. The velocity of the lower wall is kept at rest and the upper surface is initially set in motion with a velocity $V_0 = 1 \: m \: s^{-1}$.
\begin{figure}[!ht]
\begin{center}
\includegraphics[width=4.8cm,height=4.5cm]{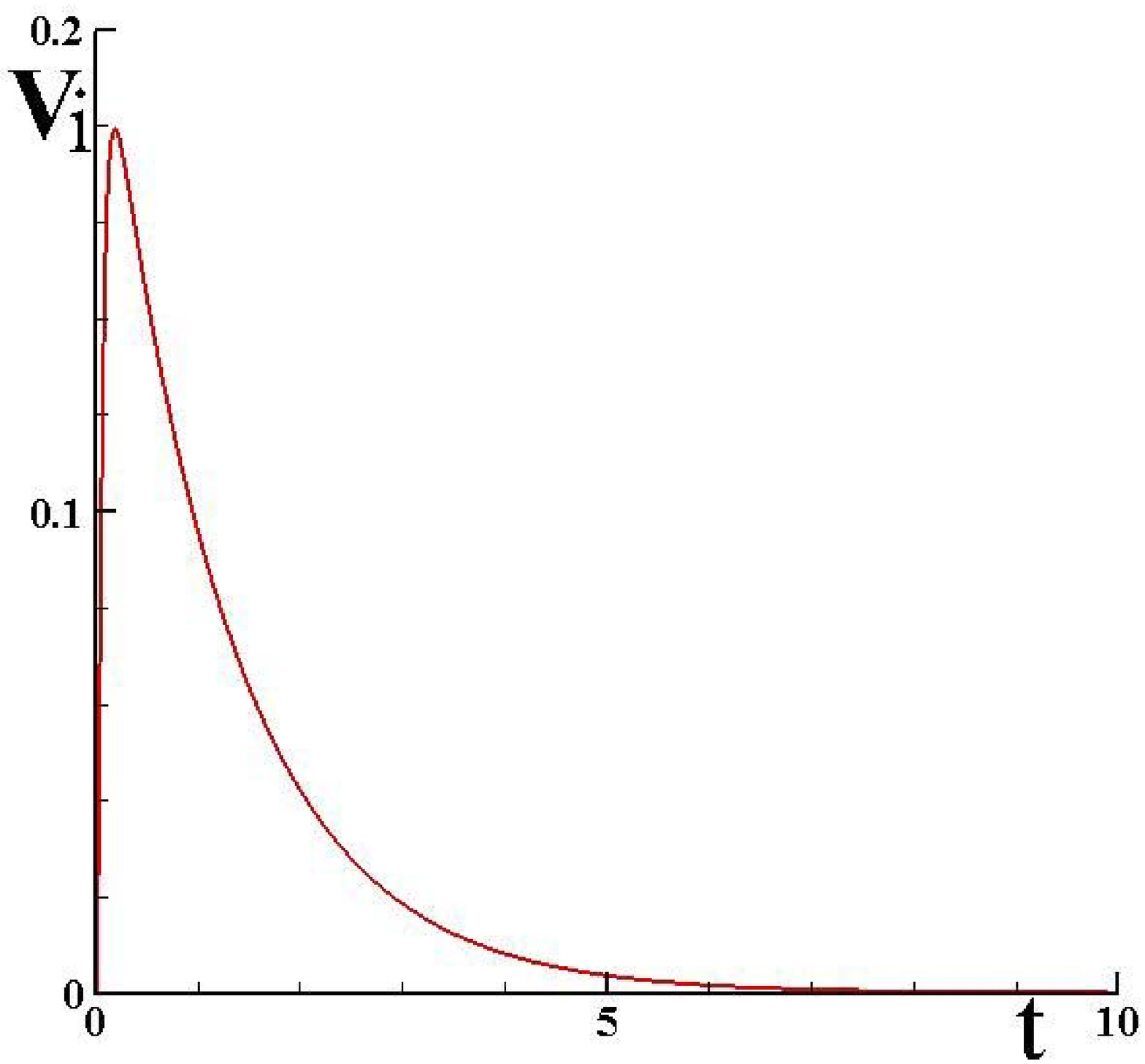}
\includegraphics[width=4.8cm,height=4.5cm]{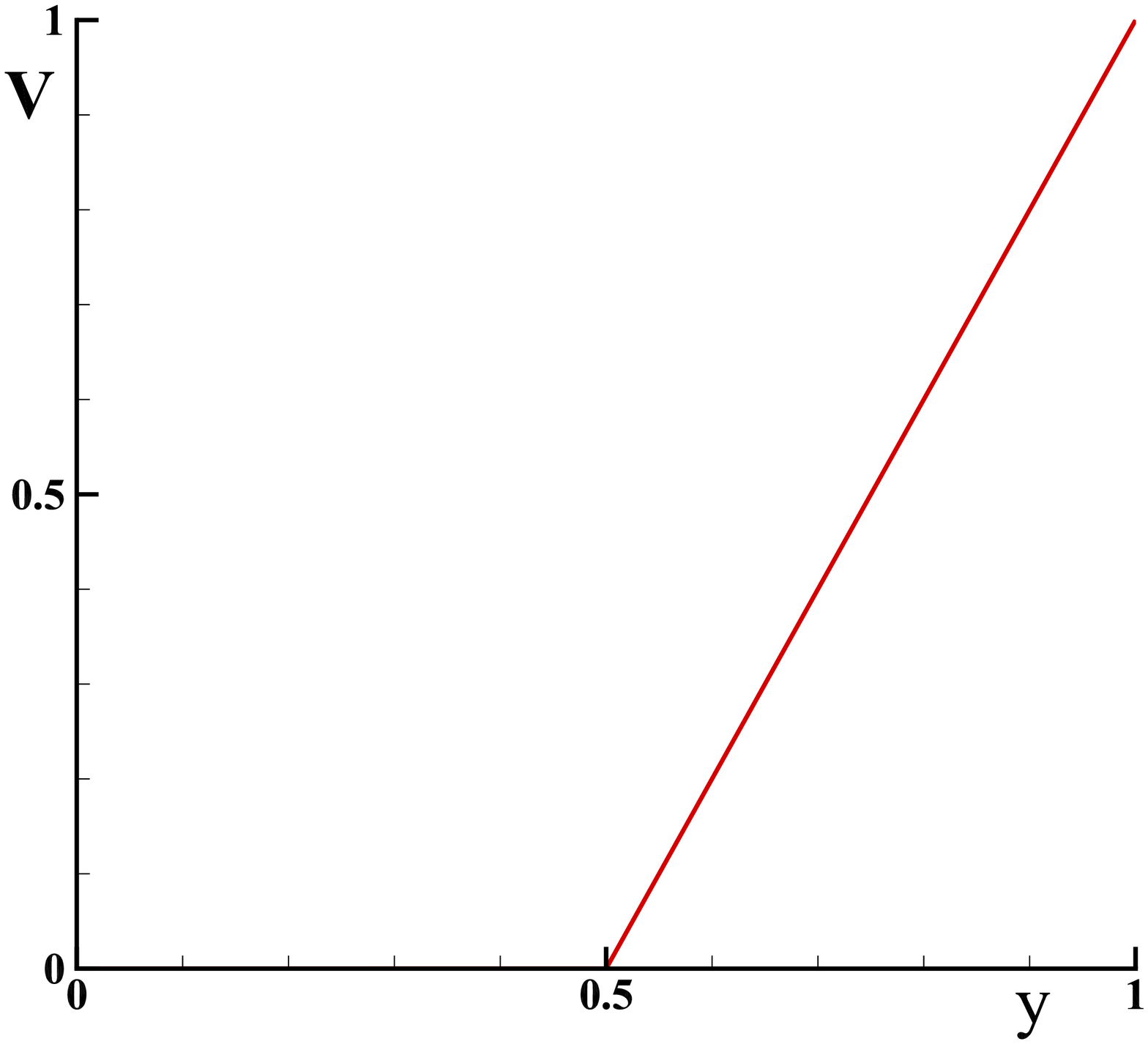}
\includegraphics[width=4.8cm,height=4.5cm]{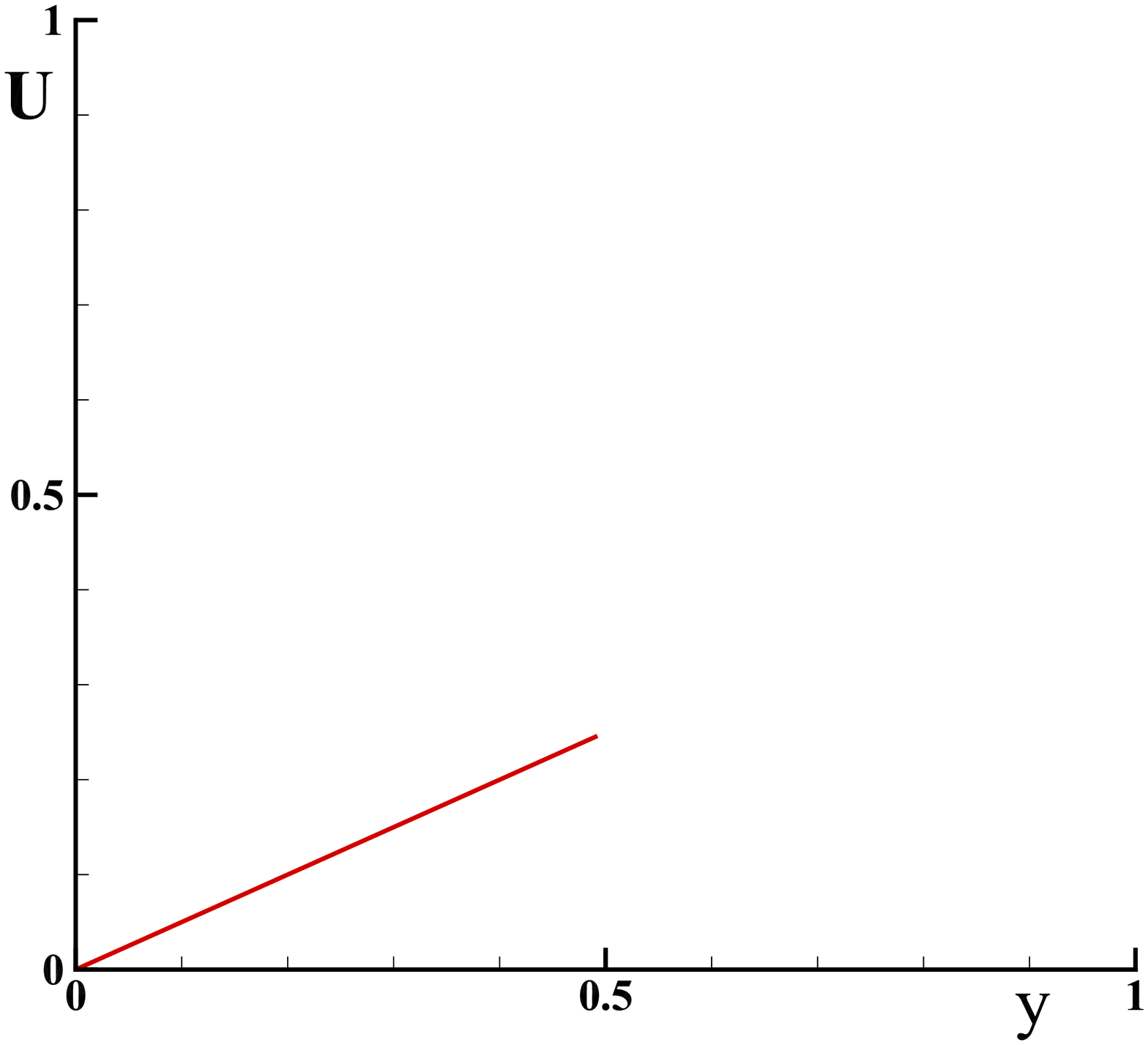}
\caption{\it Fluid-structure interaction between a viscous fluid and an elastic solid; the viscosity of the fluid is equal to $\nu = 1 \: m^2 \: s^{-1}$ and the solid shear modulus is equal to $\nu = 4 \: m^2 \: s^{-1}$. On the left, the velocity of the interface over time is presented, in the center, the velocity $\mathbf V$ at steady-state regime is reported and on the right, the displacement of the solid $\mathbf U$ plotted.}
\label{elas}
\end{center}
\end{figure}

Let us first consider the purely viscous case of two kinematic viscosity fluids $\nu_1 = 1 \: m^2 \: s^{-1}$ and $\nu_2 = 4 \: m^2 \: s^{-1}$; the solution obtained using the system (\ref{discrete}) converges very quickly towards the stationary solution. It appears as two right-hand portions satisfying the boundary conditions and the condition $\nu_1 \: \nabla \times \mathbf V_1 = \nu_2 \: \nabla \times \mathbf V_2$ at the interface, since the density on $\Sigma$ is unique. Under these conditions, the velocity at the interface is equal to $V_i = 0.2$. The 1D solution does not depend on the chosen spatial approximation and the error is zero to almost machine accuracy. Note that the condition at the interface is implicitly provided by the $\nabla \times \left (\nu \: \nabla \times \mathbf V \right)$ operator. The constant rotational in each medium is respectively equal to $\nabla \times \mathbf V_1 = -1.6$ and $\nabla \times \mathbf V_2 = -0.4$. Since the problem has no compressibility terms, only the viscous terms independent of the first ones appear in the discrete motion equation.

The lower part of the domain is now assumed to behave as an elastic solid of celerity $c_t^2 = \nu = 4$. The upper part is occupied by a fluid  whose viscosity is equal to $\nu = 1$. The potential vector $\bm \psi^o$ makes it possible to accumulate the shear stresses in the solid, the constraints at the interface in the fluid being effectively transmitted and stored in the solid. The solution converges rapidly to a strictly zero velocity field in the solid and a linear velocity profile satisfying the condition in $y = h$ and at zero velocity on the $\Sigma$ interface. The vector equation of the system (\ref{discrete}) is identically satisfied with $\bm \psi^o = \nu \: \nabla \times \mathbf V$ where $\mathbf V$ is the velocity of the fluid and $\bm \psi^o =  2$. The exact solution does not depend on the spatial approximation.

Figure (\ref{elas}) shows the evolution of the velocity at interface $\Sigma$ over time. It diminishes quickly, enough to become zero. The velocity field is zero in the solid and linear in the steady-state fluid. The figure also gives the displacement $\mathbf U$ of the solid at the end of the time evolution.

While a fluid moves indefinitely under the action of shear, an elastic solid quickly reaches a stationary displacement. The absence of interpolation at the interface between a fluid and a solid allows us to reach the exact solution. This very simple example makes it possible to understand the different mechanisms involved in the equation (\ref{discrete}) and to validate the unsteady and stationary fluid-solid interaction.

In continuum mechanics, the theoretical solution of this problem can be obtained by considering the two media separately by imposing boundary conditions at the interface.
The respective equations, in a stationary incompressible regime without inertial effects, are respectively for the fluid and solid media:
\begin{eqnarray}
\left\{
\begin{array}{llllll}
\displaystyle{  \nabla \cdot \left( \mu_f \: \left( \nabla \mathbf V + \nabla^t \mathbf V \right) \right) = 0 } \\  \\
\displaystyle{ \nabla \times  \left( \mu_s \: \nabla \times \mathbf U \right) = 0  }
\end{array}
\right.
\label{dissol}
\end{eqnarray}

When the properties $\mu_f$ and $\mu_s$ are constant, these equations are reduced to Laplacian terms.
With the assumptions adopted here, the results in the fluid are obtained with the Navier-Stokes equation, while the solid solutions come from the Navier-Lam{\'e} equation.
The conditions at the interface are simple, for the fluid the velocity is zero at $y = h / 2$, while its value is $V_0$ at $y = h$. For the solid, the displacement is null at $y = 0$ and the constraint is imposed at the interface $y = h$, chosen equal to that of the fluid side. The velocity is of course zero in the solid domain.
The solution is very simple: $v (y) = \mathbf V \cdot e_x = \left (2 \: y / h - 1 \right) $ and $u(y) = \mathbf U \cdot e_x = \mu_f \: / \mu_s \: (2 \: y / h)$. As expected, the velocity solution $v(y)$ does not depend on viscosity, whereas the displacement depends on the ratio $\mu_f / \mu_s$.
For this simple problem, the solutions of discrete mechanics are of course the same as in continuum mechanics. Among the advantages of the monolithic discrete approach, the equation of motion is unique for all media. Its acceleration formulation makes it possible to consider velocity and displacement as simple accumulators associated with operators $\nabla \cdot \mathbf V$ and $\nabla \times \mathbf V$.

\subsubsection{Extension to other constitutive laws } 

When the media have more complex rheologies, {i.e.} viscoelastic fluids, non-linear viscosity laws, viscoplastic fluids or time-dependent properties, it is possible to represent their behavior {\it a priori} in complex situations.
In particular, the accumulation of shear-rotation constraints can only be partial and a weighting of the accumulation term of $\bm \psi^o$ by an accumulation factor $0 \leq \alpha_t \leq$ 1 makes it possible to account for viscoelastic behavior. Threshold fluids are also easily represented by specifying the value of $\bm \psi^o = \bm \psi_c $ below which the medium behaves like an elastic solid. The rheology case with non-linear viscosities is no longer a difficulty.
In fact, discrete mechanics leads us to consider the notion of viscosity and that of shear-rotation as attached only to the faces of the primal topology where the constraint is expressed in the form $ \nu \: \nabla \times \mathbf V $ in fluids and $dt \: \nu \: \nabla \times \mathbf V$ in solids.

As an example, the interaction between an incompressible Newtonian viscous fluid and a Neo-Hookean elastic solid is now studied. The stress tensor expression of an incompressible isotropic hyperelastic material for the Neo-Hookean model is written as $\bm \sigma_s = - p \: \mathbf I + \mu_s \: \mathbf B $
where $\mathbf B = \mathbf F \: \mathbf F^t $ is the Cauchy-Green deformation tensor on the left.
In two space dimensions, Cayley-Hamilton's theorem shows that the model of Mooney-Rivlin hyperelastic material is equivalent to the Neo-Hookean model.

The case published by K. Sugiyama in 2011 \cite{Sug11} presents a problem relating to an elastic band solicited in shear by an incompressible Newtonian fluid flow periodic in time. The laminar flow is periodic along $ x $. In the absence of the inertia terms, the problem can be solved in one dimension of space along direction $y$, with $ y \in [0, 1] $. In the present configuration, the upper interface is animated by a periodic motion $ V(t) = V_0 \: \sin (\omega \: t) $ with $V_0 = 1 $ and $ \omega = \pi $ and the lower surface is maintained at zero velocity.
\begin{figure}[!ht]
\begin{center}
\includegraphics[width=4.3cm,height=4.cm]{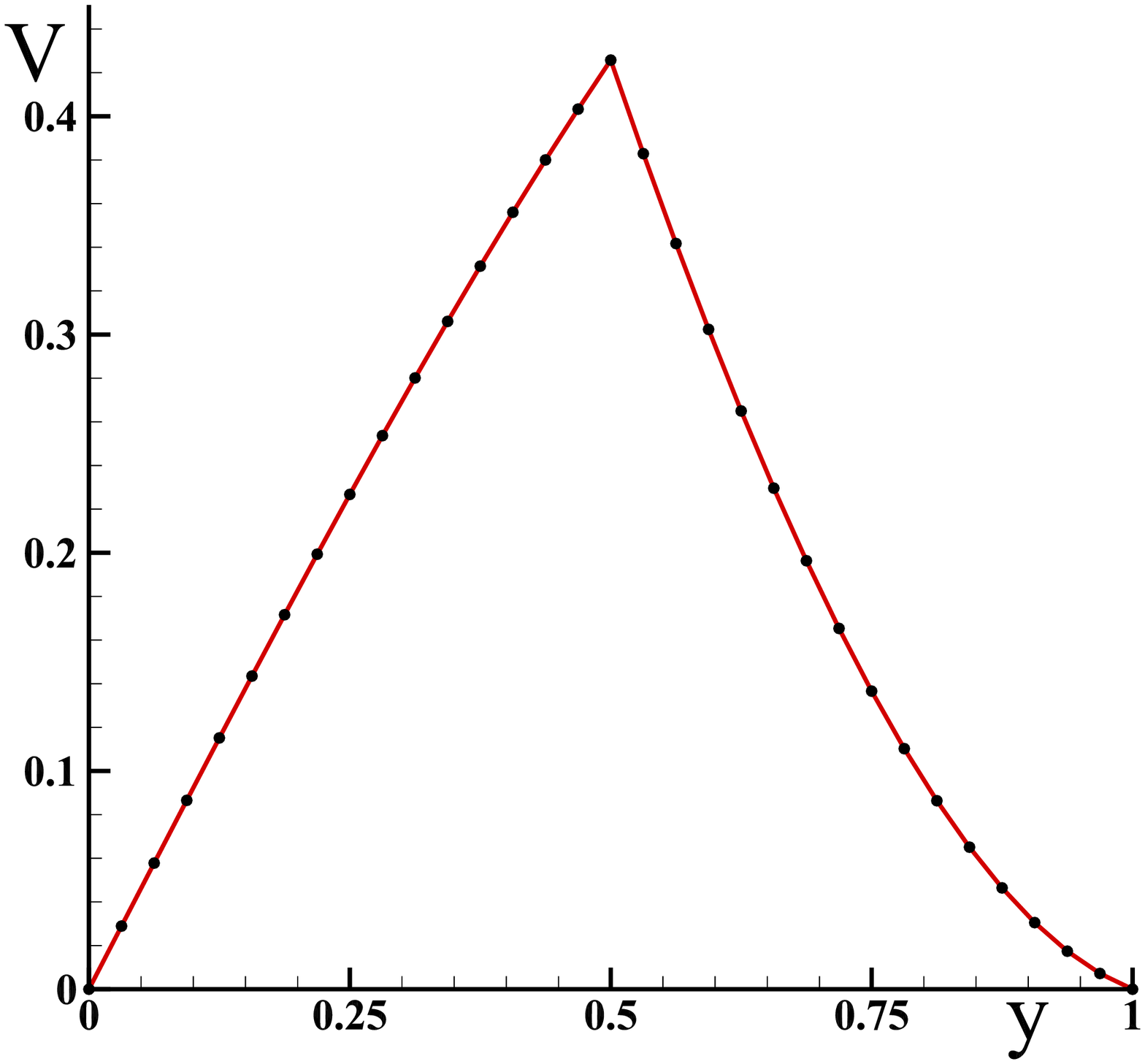}
\includegraphics[width=4.3cm,height=4.cm]{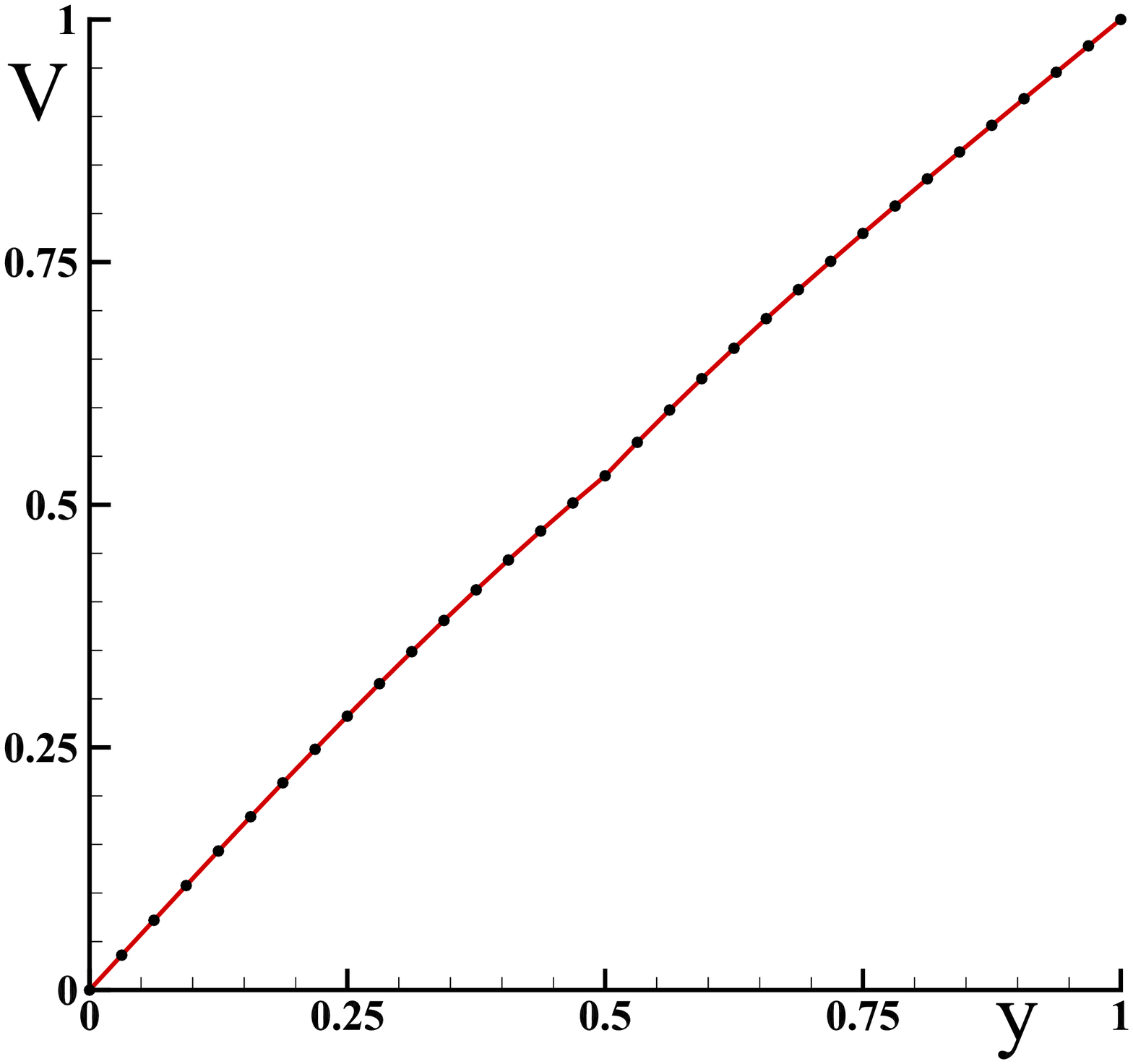}
\includegraphics[width=4.3cm,height=4.cm]{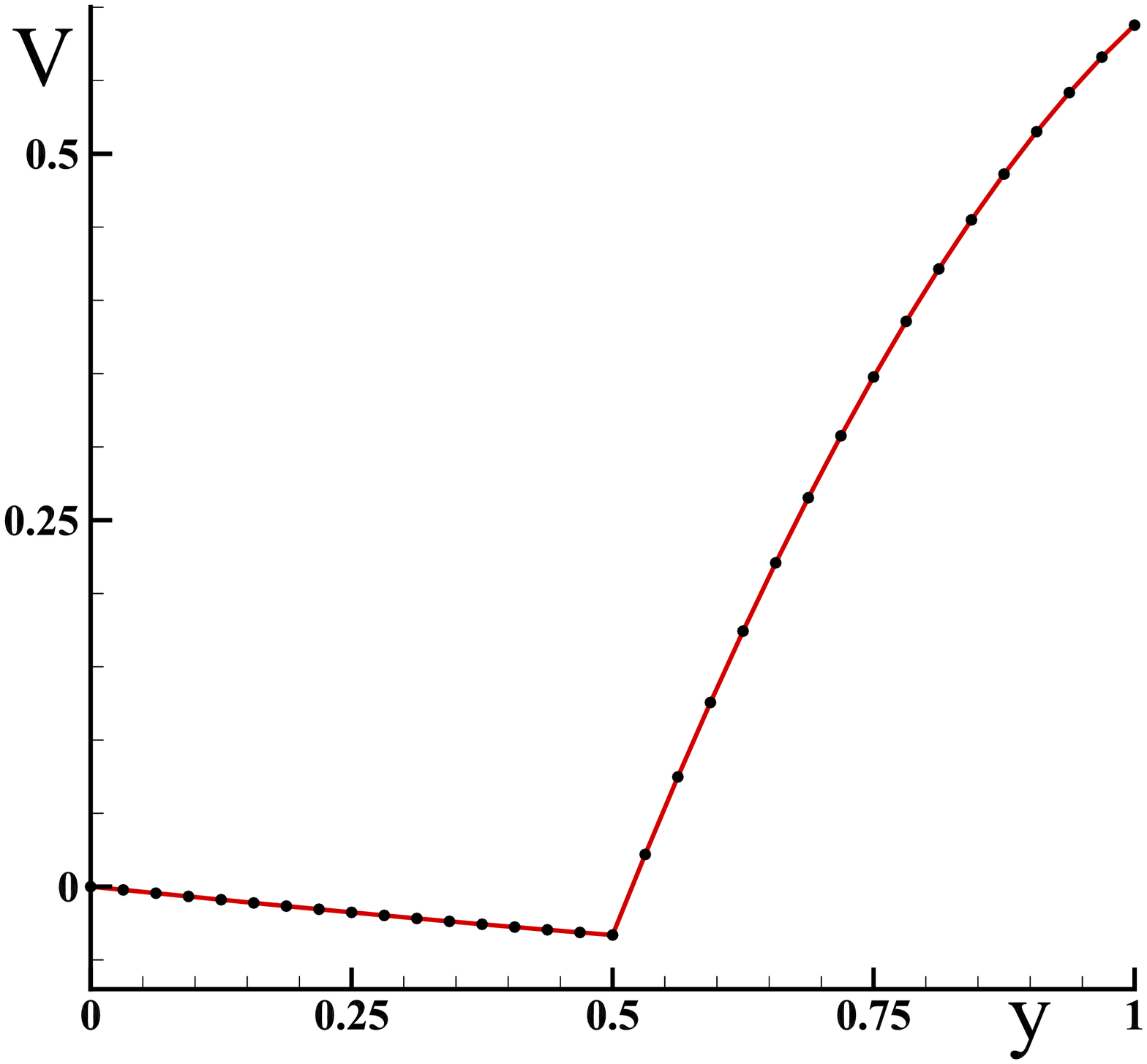}
\caption{\it Study of the fluid-structure interaction between a viscous fluid and an elastic solid with a periodic evolution. The viscosity of the fluid is equal to $\nu = 1 \: m^2 \: s^{-1}$ and the equivalent solid shear modulus is equal to $dt \: c_t^2 = 4 \: m^2 \: s^{-1}$. Velocity profiles based on $y$ for times $t = 10 \: s$, $t = 10.5 \: s$, $t = 10.8 \: s$ are plotted. the solid lines show the theoretical solution, while the dots correspond to a spatial approximation of a $32$ mesh for $y \in [0,1]$.}
\label{sugiyamavts}
\end{center}
\end{figure}

The solid occupies the lower part of the domain while the fluid lies in the upper part, the interface between them being located at $ y =1/2$.
The theoretical solution obtained by K. Sugiyama is based on the method of variable separation applied to $y$ and time $t$.
A homogeneous solution is sought by a development based on Fourier functions in the interval $ y \in [0,1] $ and time exponential functions separately for each of the fluid and solid domains. The boundary conditions make it possible to determine the set of Fourier coefficients by expressing the continuity of the velocities and the stresses at the interface.
The solution $V(y, t)$ is obtained directly by the equations of the discrete mechanics (\ref{discrete}) for which only the conditions in $y = 0$ and $y = 1$ are imposed. Interfacial boundary conditions enforcing continuity of velocity and constraint are implicitly verified by the dual curl operator. The notion of 2D or 3D space does not exist in discrete mechanics, the operators orient the normal and tangential directions in a three-dimensional space. Despite this, in the present case, depending on the adopted assumptions, the resolution is performed in one space dimension. The chosen time step is equal to $\delta t = 10^{-4} \: s$ in order to guarantee good overall accuracy. From the knowledge of the theoretical solution, the numerical solution is shown to behave with order two in space and time.
 
The solution is established very quickly. Some periods are necessary to obtain a periodic evolution of the velocity and the profiles of the velocity are recorded starting from a time $t = 10 \: s$. The displacement of the solid is computed over time by the relation $\mathbf U = \mathbf U^o + \mathbf V \: dt$ where $dt$ is both the differential element and the increment in time $\delta t = dt$.

Some profiles of the velocities following $y$ are given in figure (\ref{sugiyamavts}) as soon as the periodic regime is well established. A spatial and temporal convergence of order $2$ is observed. Given the absolute precision obtained (of the order of $ 10^{-4}$) with a coarse mesh $(n = 32)$, the error is unobservable in the comparison between theoretical solution and numerical simulation.

The case of fluid-structure interaction proposed by K. Sugiyama for a Neo-Hookean model has the advantage of providing a theoretical solution allowing a precise validation of the numerical solutions. It also provides a basis for the development of new concepts like discrete mechanics. In his publication, Sugiyama gets an error in norm $L_2$ and norm $L_{\infty}$ converging with order $1$ in space, whereas the DM model (\ref{discrete}) makes it possible to reach order two with a much lower absolute error.
This good result is due to the separation of the properties at the interface and the absence of any interpolation, in spite of an entirely monolithic and implicit treatment of the fluid-solid coupling.

Other, more complex behavior laws can be taken into account. Despite the intrinsic interest of specific studies in this area, they would not bring additional elements for the validation of the discrete model. The total disconnection between the motion equations, constitutive and state laws allows us {\it a priori} to consider, as for multiphase flows for example, the use of constitutive laws of any kind.

\subsection{Electromagnetic Fields} 

\subsubsection{Magnetic field created by infinite length wire}

This very simple case corresponds to a stationary phenomenon resulting from magnetostatics: a current $I$ runs through an electrical conductor of infinite length and very small radius; it has an electrical conductivity $\sigma$, a density $\rho$ and the permeability of the external medium is equal to that of the vacuum $\mu_0$. The degeneracy of the equation of motion (\ref{discrete}) makes it possible to obtain the equation of magnetostatics in terms of potentials:
\begin{eqnarray}
\displaystyle{ - \nabla \phi + \nabla \times \bm \psi = 0  } 
\label{fil}
\end{eqnarray}

The two quantities $\phi(x)$ and $\bm \psi(r)$ are functions of different variables and the two fields of equation (\ref{fil}) are orthogonal. The Stokes theorem and the fundamental theorem of the integral mean value make it possible to write:
\begin{eqnarray}
\displaystyle{ \int_0^{2 \pi}  \bm \psi \cdot \mathbf t \: dl = \int_a^b \nabla \phi \: dx  } 
\label{fil2}
\end{eqnarray}

To within a constant, null in this case since the lines of the magnetic field are closed, the solution of this problem is thus:
\begin{eqnarray}
\displaystyle{  \bm \psi \cdot \mathbf n  = \frac{ \left( \phi_b - \phi_a \right) }{2 \: \pi \: r} } 
\label{fil3}
\end{eqnarray}

By replacing the potentials with the usual variables of electromagnetism, $\phi = \rho_m \: e / \rho$ and $B = \rho \: \sigma \: \mu / \rho_m$ and noting that $ ( e_b - e_a) = I / \sigma$ we find the result obtained classically by the law of Biot and Savart in the form of the following component $\mathbf n$ of the magnetic field:
\begin{eqnarray}
\displaystyle{  B(r)   = \frac{ \mu_0 \: I }{2 \: \pi \: r} } 
\label{fil4}
\end{eqnarray}

The considerable interest of equation (\ref{fil}) and its solution (\ref{fil3}) bearing on the two potentials $\phi$ and $\bm \psi$, is that it is expressed only with two fundamental units, time and space. In electromagnetism, the equation and its solutions involve the other fundamental units, mass M and intensity A.

\subsubsection{Magnetic field in a torus} 

Many cases of practical interest are inspired by the design of electric motors and, in general, machines where magnetic fields and electric fields interact. The case treated here does not refer to an industrial problem, it highlights the properties of the discrete equation on a simple problem: a conductive coil is considered, made of a copper toroid, traversed by an electric current $\mathbf I$ inducing a magnetic field $\mathbf B$ in the exterior medium. The near field can be obtained by integrating the equations of the magnetostatic physics $B = \mu \: I \: R^2 / (2 (R^2 + z^2)^(3/2))$ where $R$ is the radius of the turn and $z$ is the coordinate orthogonal to the surface of the torus.
The problem is simulated by assuming that the turn is contained in a torus of elliptical section delimiting a zero electric field surface. The outer torus with elliptical cross section of $a = 1.76 \: m$ and $b = 1.2 \: m$ and the dimensions of the internal torus are $R = 1 \: $ and $d = 0.1 \: m$.

\begin{figure}[!ht]
\begin{center}
\includegraphics[width=6.cm,height=4.cm]{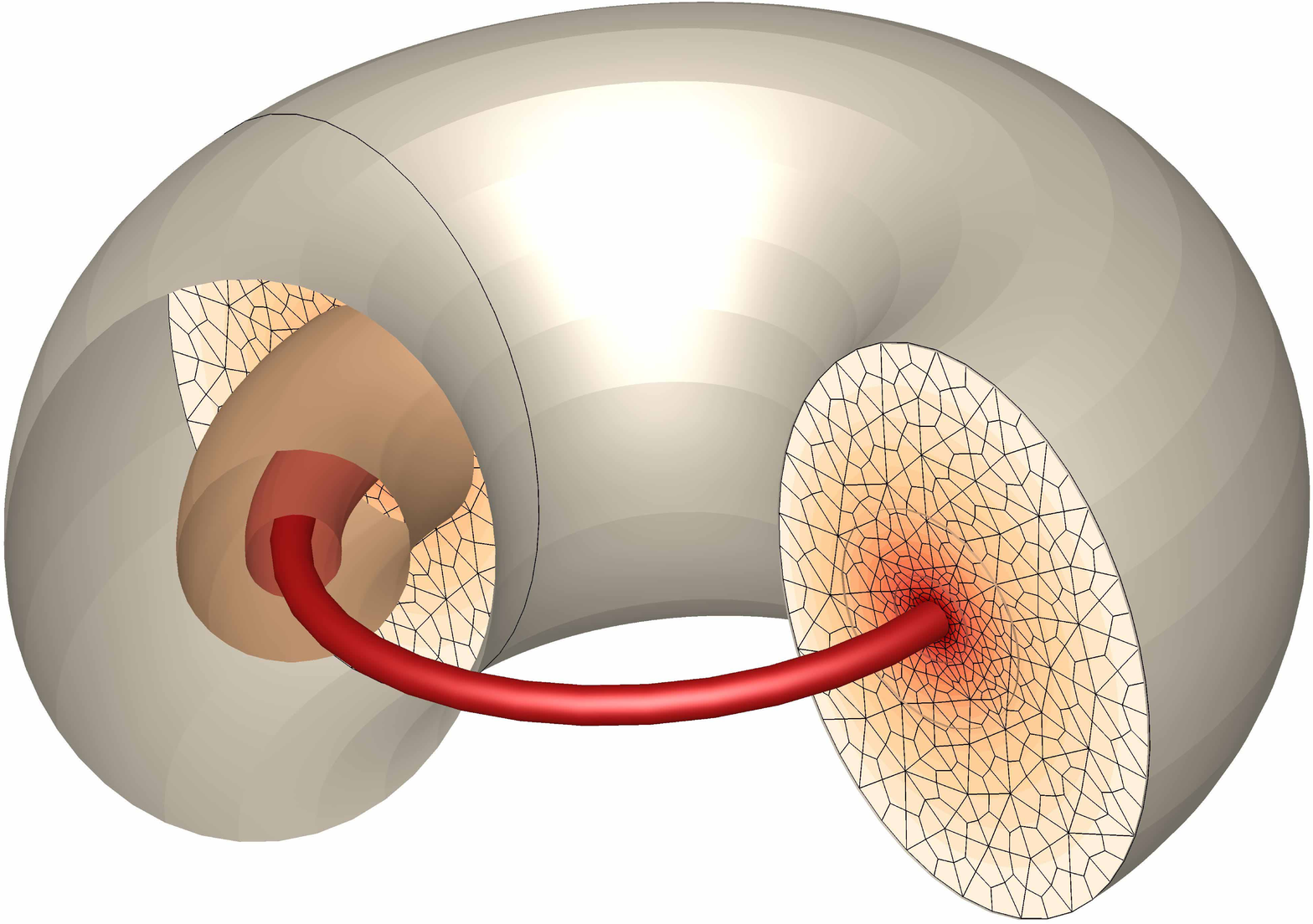} 
\includegraphics[width=3.cm,height=4.cm]{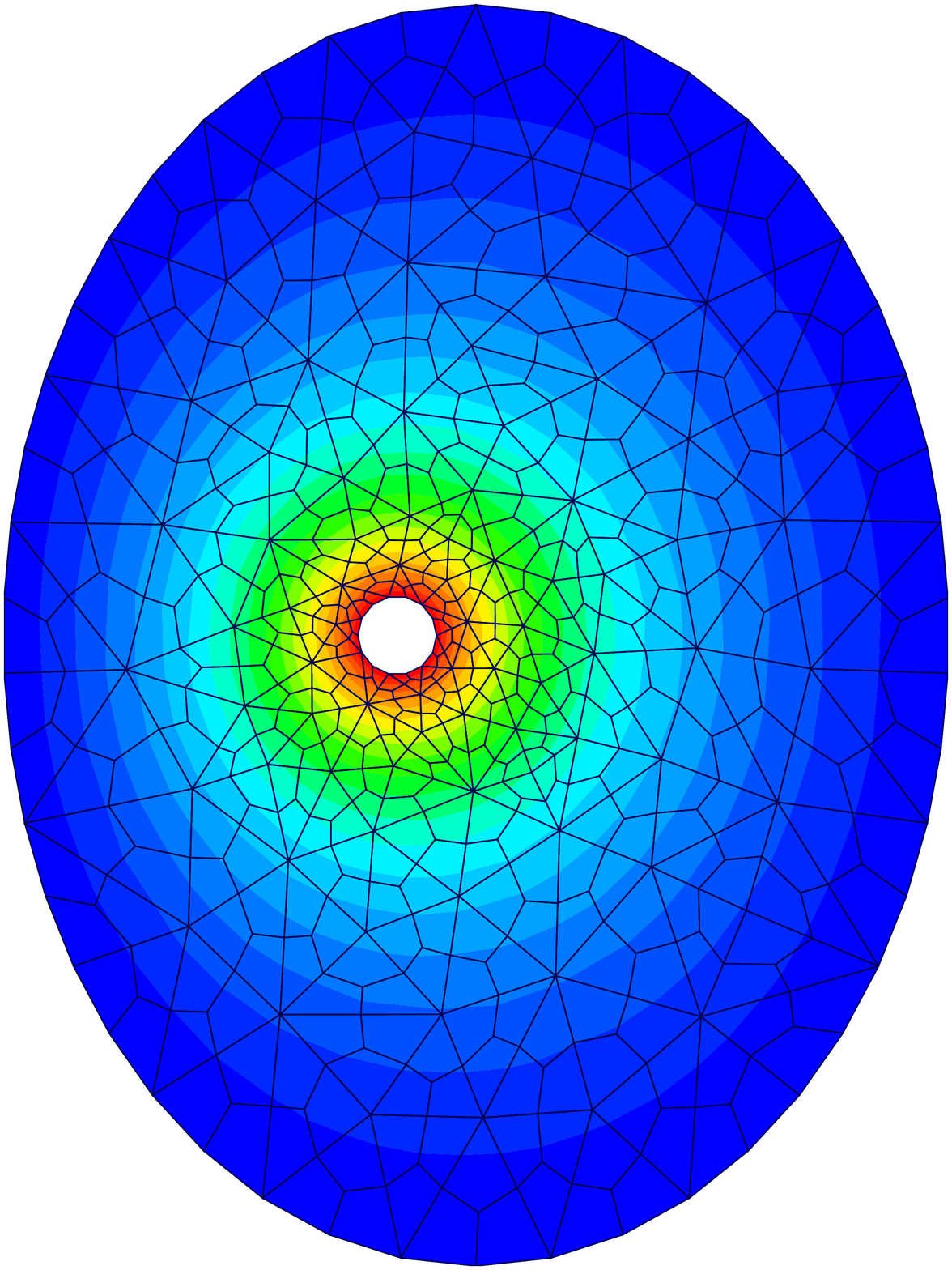} 
\vspace{-1.mm}
\caption{\it Magnetic field created by a circular turn traversed by a current $I$ within a torus of elliptical section. The unstructured mesh is composed of non-regular hexahedra \cite {Guez09}. On the left is the torus coil surrounded by isovalues of the magnetic field and on the right, a cross-section representing the potential field is shown. }
\label{tomawak}
\end{center}
\end{figure}

The three-dimensional domain is meshed with {\it gmsh} \cite{Guez09} in the form of an unstructured tessellation with a reduced number of cells conforming to the toric surfaces. Figure (\ref{tomawak}) illustrates the geometry used and the primal geometric topology, an unstructured mesh composed by $n = 120626$ hexaedra.

  From the values of the electrical potential $e$ at the ends of the turn, the properties of the media and the geometry of the chosen case, it is possible to define the scalar potential $\phi$ and to deduce the current density $\bm j$ to be imposed, represented by the velocity $\mathbf V$ in the discrete model. As the current is imposed, the resulting gradient of scalar potential is a constant $\nabla \phi^o = Cte$ and the equation to be solved becomes:
\begin{eqnarray}
\left\{
\begin{array}{llllll}
\displaystyle{ \frac{\partial \mathbf V}{\partial t} + \nabla  \left( \frac{| \mathbf V |^2 }{2}  \right) - \nabla \times \left( \frac{| \mathbf V |^2 }{2} \: \mathbf n \right)  + \nabla \times \left( \nu_m \: \nabla \times \mathbf V \right) = - \nabla  \phi^o  } \\  \\
\displaystyle{ \bm \psi  =  - \nu_m \: \nabla \times \mathbf V }
\end{array}
\right.
\label{tomawac}
\end{eqnarray}

At the end of the unsteady process, the obtained solution relates to the sole component $\mathbf V$ of the velocity on each of the edges $\Gamma$ of the primal topology. The magnetic field $\bm \psi$ is obtained at every instant by an upgrade from $\nabla \times \mathbf V$. The magnetic field lines are almost circular. They are contained in each elliptical section orthogonal to its main axis. Figure (\ref{tomawak}) shows the electric flux tubes around the coil. It should be noted that the electric field exists in the whole field, in the copper turn as well as within the elliptical core. On the contrary, the electric charge density exists only in the turn. This is actually the $\mathbf V = \bm j / \rho_m$ velocity that is being searched for and this ratio always keeps a physical meaning, even at the limit when $\rho_m \rightarrow 0$ in the absence of current.

If we neglect the inertial effects, the solution reaches a steady state for which $\nabla \times (\nu_m \: \nabla \times \mathbf V) = - \nabla \phi^o$. We consider that $\nu_m$ is a constant and that the field $\mathbf V$ is solenoidal. Under these conditions, the equation (\ref{tomawac}) becomes a simple vectorial Poisson equation $\nu_m \: \nabla^2 \bm \psi = \nabla \phi^o$. For this example, the scalar field $\psi = \bm \psi \cdot \mathbf n$ depends on two space variables and so is the same whatever the surface defined by an elliptical cross section of the external torus.

This simulation of an induced magnetic field problem deals with a stationary case with obvious symmetries but it shows the versatility of the equation of discrete motion. The problem has been solved with an unsteady formulation of the vectorial equation (\ref{tomawac}) and the solution deals with the quantities $(\mathbf V, \phi, \bm \psi)$. Since the relations between these quantities and the classical variables of electromagnetism are bijective, it is of course possible to recover these variables, even if it is not a necessity. In fact, each field of physics has its variables, its physical properties, but they are not all independent. The discrete formulation presented here not only unifies mechanics and electromagnetism, but also proposes unique variables for these domains. These variables, which are expressed only with two fundamental units, as well as the physical properties involved in the formulation, could possibly be extended to other areas of physics.

\subsection{Propagation of Light}

Two applications are treated here, the case of interference produced by a coherent light source and the refraction of a beam of light at the free surface between air and water. These phenomena taught in elementary physics courses are an opportunity to test the law (\ref{discrete}) on observations made for centuries. The modeling of these observations does not require complex theories, but the objective here is to use this law without modification by including all the terms, including the inertial effects not present in undulatory optics.

The cases treated here are classical phenomena described by the Maxwell equations and whose solutions can be obtained by analytical methods based on the resolution of a Helmholtz equation; the goal here is to use the system directly (\ref{discrete}).
In order to verify the robustness of the law (\ref{discrete}), it will not be scaled. The times and lengths are characteristic of visible light and the calculations will be done in direct simulation.

\subsubsection{Interferences produced by two coherent point sources} 

Rays of visible light, such as acoustic waves produced by the same source and traveling on different paths, induce interfering fringes materialized by alternately light and dark bands for which the distances which separate them are easily measurable. This phenomenon is one of those that helped understand and model the wave nature of light.

Let us consider two monochromatic synchronous sources of frequency $f$ placed in the planar surface $(x, y)$ in $x_i = \pm d$ and $y = 0$. The areas of equal phase difference, the places of the points $P$ for which the walking difference $\delta = | x_1 P - x_2 P |$ and the phase difference $\varphi = 2 \: \pi \: \delta / \lambda$ are constant, are hyperbolas with $x_i$ for foci. The surfaces for which $\delta = k \: \lambda$ $(k \in N)$ are ventral surfaces and the surfaces for which $\delta = (2 \: k + 1) \: (2 \: \lambda)$ are nodal surfaces.
\begin{figure}[!ht]
\begin{center}
\includegraphics[width=5.cm,height=5.cm]{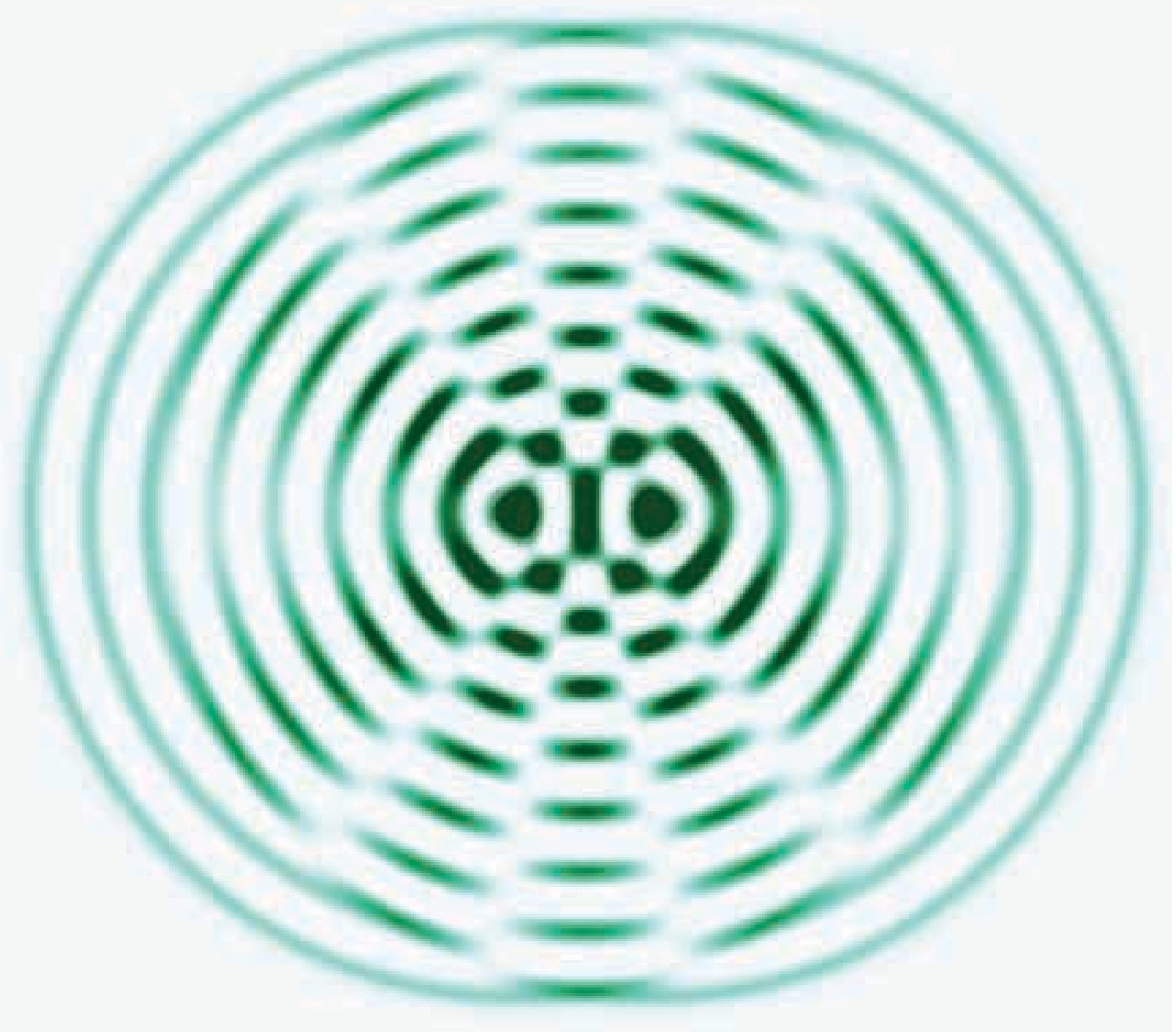}
\hspace{10.mm}
\includegraphics[width=5.cm,height=5.cm]{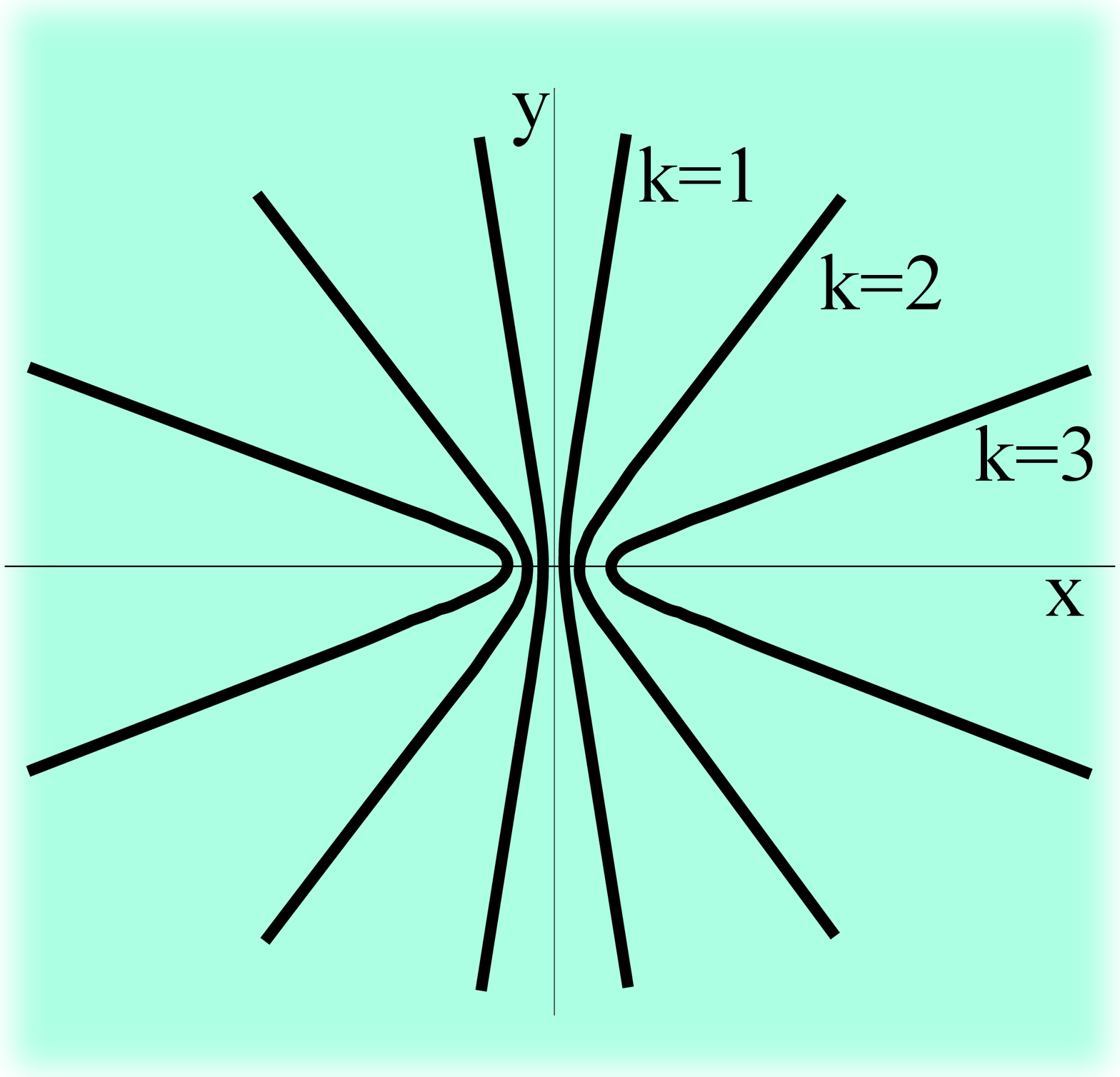} \\
\hspace{-0.mm} (a) \hspace{54.mm} (b)
\caption{\it Stationary interference fringes produced by two coherent point sources: (a) instantaneous wave field obtained by a direct simulation for a time $t = 1.4 \: 10^{-14} \: s$ (b) perception by the human eye of dark fringes.}
\label{doublet}
\end{center}
\end{figure}

The purpose of this section is to show the results of an unsteady direct simulation of the ignition, from $t = t^o$, of two synchronous sources of visible light of wavelength $\lambda$ and of frequency $f$ such that $f = c_o / \lambda$. The simulation was performed without upscaling from the system of complete equations (\ref{discrete}) including the inertial terms. The light is of wavelength $\lambda = 0.5 \: 10^{-6} \: m$, the velocity of the medium (or of the vacuum) is taken equal to $c_0 = 3 \: 10^8 \: m s^{- 1}$ and the frequency equal to $f = 6 \: 10^{14} \: s^{-1}$; the two distant sources of $ 2 \cdot d = 2 \cdot 6 \: 10^{-7} \: m$ are of radius $r = 4 \: 10^{-8} \: m$ so as to approximate them by a point $(r << d)$. The chosen time step is $ dt = 10^{-18} \: s $ and the Cartesian mesh is composed of $10^6$ cells.

From the initial time at $t = t_o$, the sources emit perfectly cylindrical progressive waves which begin to interfere for a time equal to $t_1 = d / c_0$ ie. $t_1 = 2 \: 10^{-15} \: s$. The interferences then amplify in all the domain forming a series of stationary fringes which are the places of the nodes where the amplitude remains null. These surfaces appear in a surface as dark lines to an observer who does not distinguish the very fast temporal variations of the amplitudes of the ventral surfaces. In three dimensions of space, these surfaces become hyperboloid. The increase of the distance between the two sources $2 \cdot d$, keeping the same characteristics as those fixed, would make it possible to multiply the fringes of interference between them.

The figure (\ref{doublet}) shows the scalar potential field $\phi^o$. The colored lines are progressive waves of wavelength $\lambda$; over time, they form fringes of stationary hyperbolic interference. The result is conventionally obtained by other techniques, especially from complex potentials. Here, the unsteady direct simulation is carried out from null fields. The system (\ref{discrete}) is integrated without upscaling.

\subsubsection{Refraction of a polarized monochromatic wave } 

When a wave passes through an interface separating two media of different refractive indices, the light is deflected if the ray is not normal at the interface and the phenomenon is called refraction. If we call $\theta_i$ and $\theta_t$ the angles of the incident wave and the refracted wave with the normal and $n_i$ and $n_t$ the absolute refractive indices, we have the Snell-Descartes relation:
\begin{eqnarray}
\displaystyle{  n_i \: sin \theta_i =  n_t \: sin \theta_t }
\label{refract-a}
\end{eqnarray}

The absolute indices are the ratios of the velocity of the medium considered to the celerity in the vacuum $ n = c / c_0$ according to the Huygens-Fresnel principle. This result of geometrical optics is the same in physical optics where the wavefront is deflected. This phenomenon is found for any electromagnetic wave described by the theory of J.C. Maxwell.

The case presented is that of an incident wave polarized through an air/water interface under the angle of Brewster, the angle for which the reflected wave and the transmitted wave are orthogonal; for an interface between air and water ($ n = 4/3 $) we get $ tg \theta_B = n_2 / n_1$ that is $\theta_B = 53.13 $\textsuperscript{o} and $\theta_t = 22.02$\textsuperscript{o}. The length of the interface is equal to $l = 10^{-6} \: m$, the wavelength of the incident radiation at $\lambda = 0.25 \: 10^{-6} \: m $ (ultraviolet) is a frequency of $f = 1.2 \: 10^{15} Hz $. The time-elapse $dt$ of observation is equal to $dt = 2 \: 10^{-19} s = 0.2 \: as$.

\begin{figure}[!ht]
\begin{center}
\includegraphics[width=5.cm,height=6.cm]{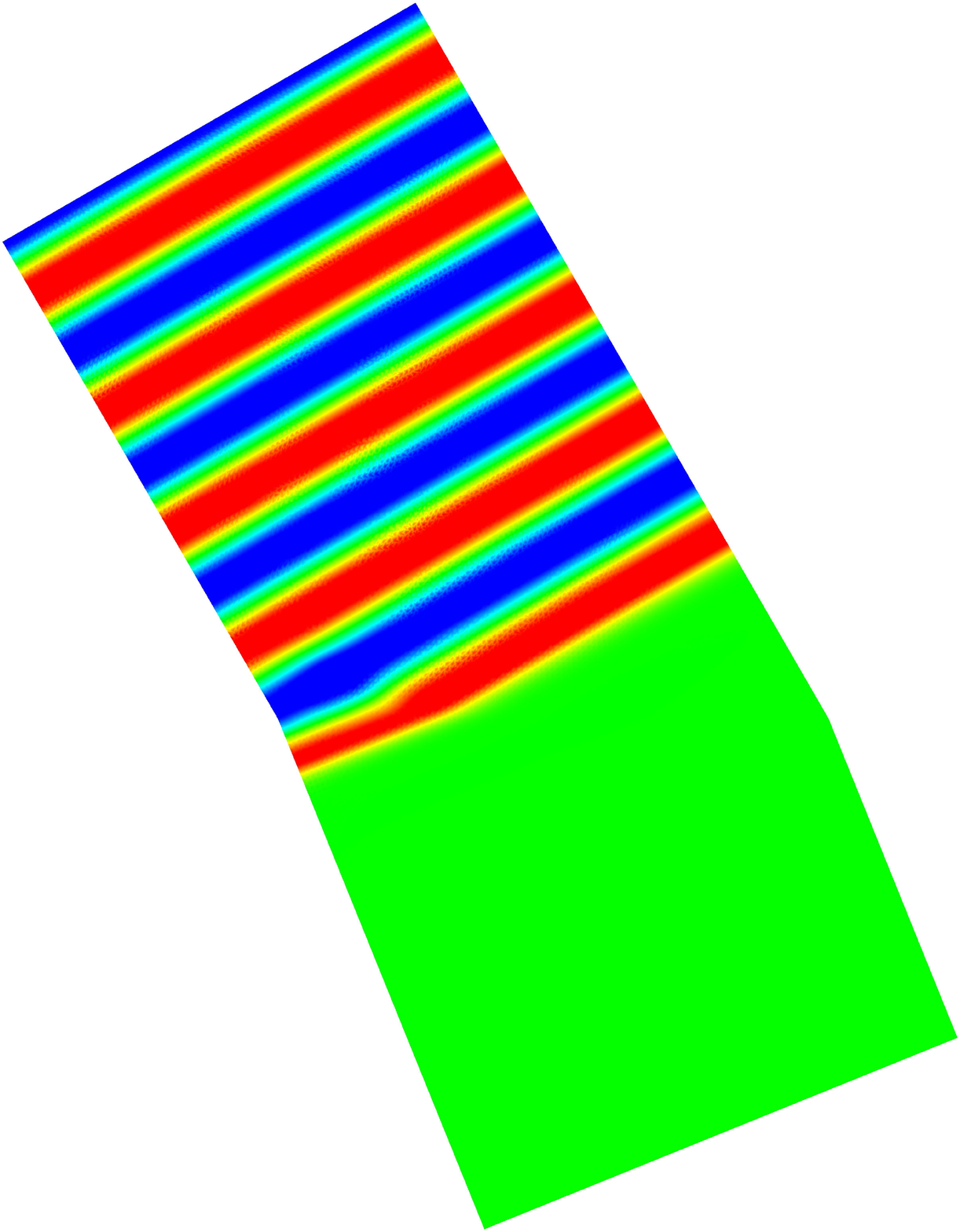}
\hspace{10.mm}
\includegraphics[width=5.cm,height=6.cm]{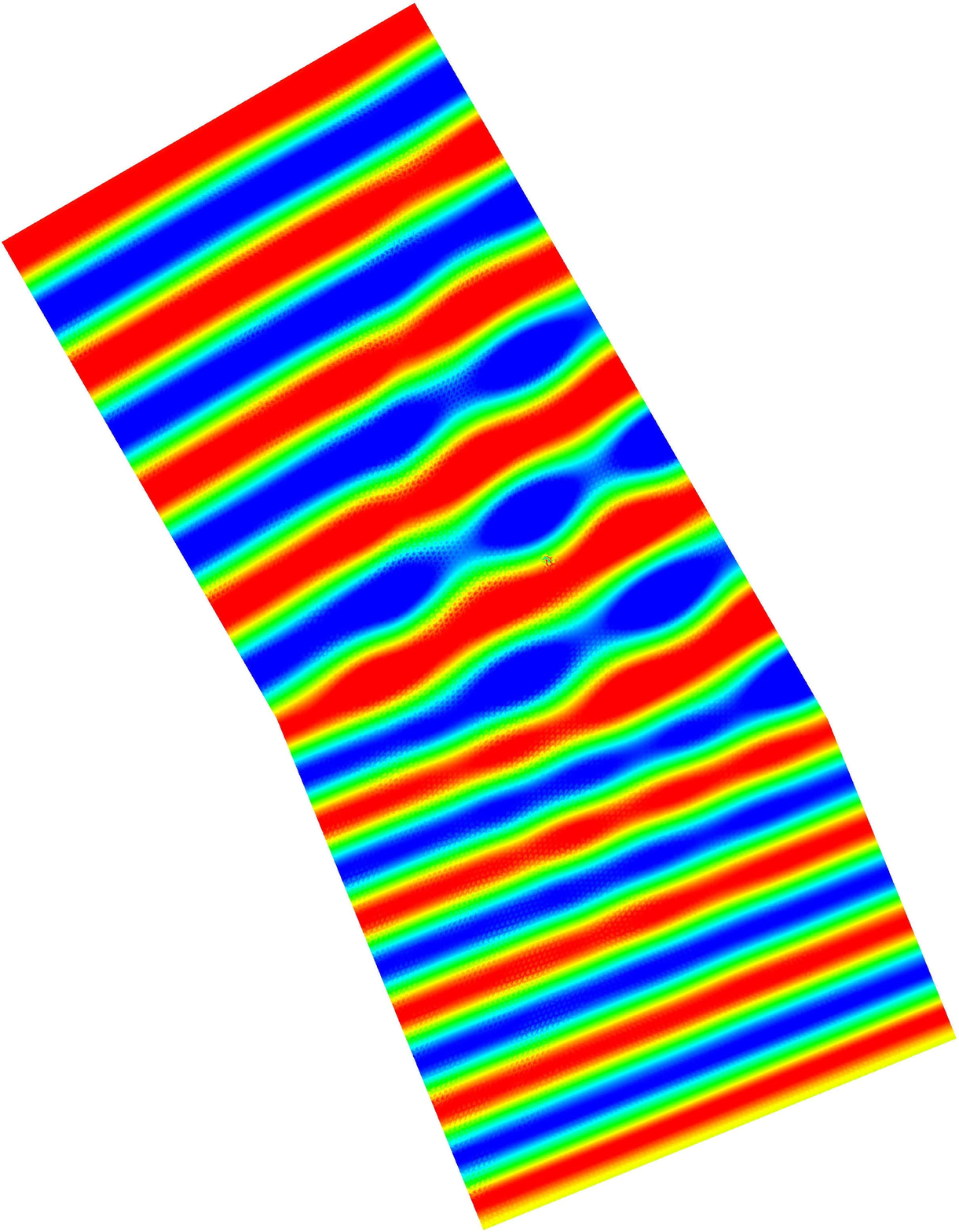}
\hspace{50.mm} (a) \hspace{54.mm} (b)
\vspace{-3.mm}
\caption{\it Refraction at the free horizontal interface between air and water of refractive index $n = 4/3$ for an incident ray inclined at $53.13$\textsuperscript{o} and a wavelength of light equal to $\lambda = 0.25 \: 10^{-6} \: m$; scalar potential $\phi^ o$ for (a) $t_1 = 1.95 \: 10^{-15} \: s$ and (b) $t_2 = 3.9 \: 10^{-15} \: s$. }
\label{refract-b}
\end{center}
\end{figure}

The figure (\ref{refract-b}) illustrates the progression of the incident wave through the air/water interface for two different times $t_1 = 1.95 \: 10^{-15} \: s$ and (b ) $t_2 = 3.9 \: 10^{-15} \: s$. The refraction of the incident wave is observed with a decrease of the wavelength as the theory predicts. 

\vspace{5.mm}

Like the first example, this shows the ability of the discrete motion law to predict the behavior of fluids or solids subjected to compressive or shear stresses at large time constants and, without any modification, to show the propagation of the light.

\section{Conclusions} 

Mechanics and electromagnetism considered as two independent branches are in fact derived from the same physics. The equation of discrete motion is derived from the original concepts of Galileo, the principle of equivalence between inertial mass and gravitational mass and the principle of relativity. Discrete mechanics diverges rapidly from classical theories as soon as Newton's second law is pronounced by suppressing the mass from it. The acceleration of a particle or a medium becomes the sum of the accelerations it undergoes and can be considered as an absolute quantity. Each acceleration can be decomposed into a solenoidal and an irrotational component following a Hodge-Helmholtz decomposition. 

The potentials $\phi$ and $\bm \psi$ retain meaning in all media, fluid, solid or vacuum; whereas, for example, the density and the pressure no longer have any meaning in a vacuum, the ratio $\phi = p / \rho$ keeps a physical meaning. Similarly, the velocity of the vacuum that can be associated with the vacuum displacement current in electromagnetism is an indispensable notion to represent the advection phenomenon.

The main contributions reported in this article are:
\begin{itemize}
\item the equation is the same for mechanics, solids and fluids, and electromagnetism;
\item $\mathbf V$ components of velocity on $\Gamma$ edges are the only unknowns;
\item the two potentials $\phi$ and $\bm \psi$ of the acceleration are representative of a common physics;
\item all quantities, variables and properties are expressed with only two fundamental units.
\end{itemize}

Of course, the proposed discrete formulation does not invalidate the previous theories, it generalizes and unifies the two major areas of physics. The integration of absent inertial terms into Maxwell's equations is an absolute necessity, first for a potential unification, but above all for expressing a physical reality.

Even from the point of view of the numerical methodology, the contribution is not negligible either. The discrete equation is ready to use, it is enough to consider the primal and dual topologies and to build the gradient, divergence, primal and dual curl. These operators immediately satisfy the discrete properties $\nabla_h \times \nabla_h \phi = 0$ and $\nabla_h \cdot (\nabla_h \times \bm \psi) = 0$. The assembly of these discrete operators leads to an algebraic system of equations in which the unknowns are the components of the velocity. Its resolution then makes it possible to upgrade the potentials $\phi$ and $\bm \psi$.

\bibliographystyle{plain}
\bibliography{D:/calta/tex/database}

\end{document}